\documentclass[twocolumn,superscriptaddress,secnumarabic,amssymb, nobibnotes, aps, bm,pre,floatfix,longbibliography]{revtex4-2}

\usepackage{graphicx}
\usepackage[usenames,dvipsnames]{xcolor}
\usepackage{comment}
\usepackage{dcolumn}
\usepackage{bm}
\usepackage{physics}
\usepackage{xcolor}
\usepackage[colorlinks,urlcolor=goodgreen,citecolor=blue,linkcolor=goodred]{hyperref}
\usepackage{cleveref}
\usepackage{float}
\definecolor{oceanboatblue}{rgb}{0.0, 0.47, 0.75}
\definecolor{orange}{rgb}{1,0.5,0}
\definecolor{goodgreen}{rgb}{0.1,0.5,0}
\definecolor{goodred}{rgb}{0.7,0,0}
\definecolor{tangerine}{rgb}{0.95, 0.52, 0.0}

\renewcommand{\vec}[1]{\boldsymbol{#1}}

\begin{document}

\title{Probing Magnetic and Triplet Correlations in Spin-Split Superconductors\\ 
with Magnetic Impurities}
\author{C.-H. Huang}
\affiliation{Donostia International Physics Center (DIPC), 20018 Donostia--San Sebasti\'an, Spain}

\author{A. Skurativska} 
\affiliation{Donostia International Physics Center (DIPC), 20018 Donostia--San Sebasti\'an, Spain}

\author{F.~S. Bergeret}
\affiliation{Centro de F\'isica de Materiales (CFM-MPC) Centro Mixto CSIC-UPV/EHU,
20018 Donostia-San Sebasti\'an, 
Basque Country, Spain}
\affiliation{Donostia International Physics Center (DIPC), 20018 Donostia--San Sebasti\'an, Spain}

\author{M. A. Cazalilla}
\affiliation{Donostia International Physics Center (DIPC), 20018 Donostia--San Sebasti\'an, Spain}
\affiliation{IKERBASQUE, Basque Foundation for Science, Plaza Euskadi 5
48009 Bilbao, Spain}

\date{\today}
\begin{abstract}
    A superconductor (SC) in proximity to a ferromagnetic insulator (FMI) is predicted to exhibit mixed singlet and triplet pair correlations.
    The magnetic proximity effect of FMI spin-splits the energy of Bogoliubov excitations 
    and leads to a spin polarization at the surface for superconducting films
    thinner than the superconducting coherence length. In this work, we study manifestations of 
    these phenomena in the properties of a magnetic impurity coupled via Kondo coupling 
    to this FMI/SC system. Using the numerical renormalization group (NRG) method, we compute the properties of the ground state and low-lying excited states of a model that incorporates the Kondo interaction and a Ruderman-Kittel-Kasuya-Yosida (RKKY)-like interaction 
    with the surface spin polarization. Our main finding is an energy splitting of the lowest even fermion-parity states caused by the proximity to the FMI. 
    As the Kondo coupling increases, the splitting grows and saturates to a universal value equal to twice the exchange field of the FMI. We introduce a two-site model that can be solved analytically and provides a qualitative understanding of this and other NRG results. In addition, using perturbation theory we demonstrate that the mechanism behind the splitting involves the RKKY field and the triplet correlations of the spin-split superconductor. A scaling analysis combined with NRG shows that the splitting can be written as a single-parameter scaling function of the ratio of the Kondo temperature and the superconducting gap, which is also numerically obtained.
\end{abstract}

\maketitle

\section{Introduction}

The search for superconductors with spin-triplet electron pairing mechanism is an ongoing endeavor motivated by their potential to host unique excitations~\cite{kitaev2001, nadj2014observation,ruby2015tunneling} with promising applications to quantum hardware~\cite{beenakker2013,sarma2015}. Within this area, one research direction focuses on studying candidate materials where this type of pairing occurs intrinsically~\cite{mackenzie2003superconductivity, ran2019nearly, zhou2022isospin}. Alternatively, triplet correlations can be induced  \emph{extrinsically} in e.g. superconductor-ferromagnet hybrid systems~\cite{bergeret2005odd}, or  conventional superconductors in proximity to a ferromagnetic insulator ~\cite{bergeret:2018colloquium}. 
 
In a conventional superconductor, Cooper pairs condense in a spin-singlet state. When the condensate interacts with a local exchange field via proximity effect either to a metallic ferromagnet or ferromagnetic insulator, part of the singlets transform into triplet pairs. 
Theoretically, this has been extensively investigated using the quasi-classical approach in Refs.~\cite{bergeret2005odd, eschrig2008triplet,hijano2021coexistence}. In the particular case of a ferromagnetic insulator (FMI)/superconductor (SC) heterostructure, an indirect signature of the induced exchange field and triplet correlations  can be observed as a spin-splitting of the density of states of thin SC layers, even in the absence of an externally applied field, as confirmed in several experiments~\cite{Moodera:1988,meservey1994spin,strambini2017revealing}.

The ferromagnet-superconductor interaction can be described treating the magnetization of the ferromagnet as a classical variable that provides an exchange field at the boundary. On the other hand, the study of magnetic impurities in a superconducting host is a longstanding area of research, with current interest focusing on the development of experimental platforms that allow to control and tune the quantum states of such systems with high accuracy. Advances in Scanning Tunneling Spectroscopy (STS) allow to probe impurities on the surface of a superconductor with atomic-scale resolution~\cite{eigler:1997, ji:2008, heinrich:2018,franke:2021}. Molecular junctions~\cite{herre:2017} and superconducting nanowires coupled to a quantum dot~\cite{lim2015shiba,jellinggaard2016tuning,valentini2021nontopological, lee:2014, pita:2022,bargerbos:2022} can be tuned to regimes where they can be modelled as quantum impurities. 

  Furthermore, chains of magnetic impurities (adatoms) on top of a superconductor have been extensively studied to realize the paradigmatic Kitaev chain~\cite{kitaev2001, nadj2014observation, ruby2015tunneling, pawlak2016probing, jeon2017distinguishing}. Recently, two nearby magnetic impurities in a superconductor have been proposed to realize an effective two-level system, which is a key ingredient of a qubit, the building block of a quantum computer~\cite{mircea:2021}. Despite the proposals, control of the system parameters remains a major challenge, requiring a deeper understanding of the magnetic interactions between impurities and the substrate. When two impurities are placed on a superconductor their interaction is mediated by the itinerant electrons resulting in an effective long-range interaction. Depending on the distance between the impurities the Ruderman-Kittel-Kasuya-Yosida (RKKY) can be antiferromagnetic~\cite{Galitski:2002, yao:2014} or even ferromagnetic at distances larger than the coherence length in superconductors with strong spin-orbit coupling~\cite{lu2023ferromagnetic}.  Some recent proposals have also discussed tuning magnetic interactions by applying microwave fields~\cite{akkaravarawong2019}, or by varying the orientation of an external magnetic field~\cite{li2016manipulating}. 
In this regard, it is interesting to explore the possibility of manipulating the magnetic interaction between impurities through an effective exchange field generated via the  proximity to an FMI/SC system at zero external magnetic field. The first step in this direction is to study how  the exchange field affects the ground state and the spectral properties of a single impurity.

In this work, we address this question by analyzing a FMI/SC heterostructure coupled to a spin-$\frac{1}{2}$ magnetic impurity. Such a system can be  seen as a platform to study the interplay between triplet correlations and the magnetic interactions induced by the FMI. As mentioned above, a FMI in proximity to a SC leads to spin polarization of the quasi-particle states, resulting in spin-splitting of the coherence peaks in the density of states ~\cite{Moodera:1988, strambini2017revealing}. 
To capture this effect, we introduce an effective homogeneous exchange field $h$ that couples to the electronic states described by the BCS mean-field Hamiltonian as a Zeeman coupling. This is justified by assuming that the thickness of the superconducting layer is smaller than the superconducting coherence length. 
The presence of this effective $h$ modifies drastically the exchange interaction between the impurity spin and the electrons in the SC. In addition, Cooper pairs in the SC mediate an effective RKKY-like interaction between the impurity spin and the FMI. 

After introducing the model, we describe the results obtained by solving it with the numerical renormalization group (NRG) method~\cite{RevModPhys.47.773_NRG,JPSJ.67.1332_BCS_trans,JPSJ.67.1332_BCS_trans2,JPSJ.69.1812_BCS_trans3,RevModPhys.80.395_NRG} for the ground state and the low-lying energy spectrum. In the absence of the FMI, the ground state properties of a superconductor/impurity system have been studied extensively~\cite{Balatsky_2006}. It was found that,
as the (Kondo) exchange coupling with the impurity grows, the system undergoes a quantum phase transition  where the fermion parity of the ground state changes from an even-parity doublet to an odd-parity siglet state. Our main finding in this work is that the presence of an exchange field induced by proximity to the FMI lifts spin degeneracy of the even-parity doublet. The spin-splitting of the doublet exists  for any finite value of the Kondo coupling. Furthermore, in the weak coupling regime, we find a shift of the threshold for continuum excitations only for electrons tunneling with spin anti-parallel to the ground state spin. 
Previously, Ref.~\cite{zitko:2017} addressed the problem of a quantum impurity coupled to a spin-split superconductor using NRG. Similarly to this work, a splitting of the doublet states was observed. However, in the model of Ref.~\cite{zitko:2017} this effect requires an in-plane external magnetic field, whilst in our system the splitting stems from the proximity to FMI and the system parameters are different from those studied in~\cite{zitko:2017}.

 Moreover, although the results obtained using NRG are quantitatively accurate, they do not provide an intuitive picture of the underlying mechanisms leading to the splitting of the doublet and the shift of the threshold for the continuum single-particle excitations. Thus, in order to gain a qualitative understanding, we introduce a minimal two-site model based on zero-bandwidth approximation~\cite{von2021yu,vecino2003josephson} that captures the essence of the NRG results. Furthermore, in the limit of weak Kondo coupling, we use perturbation theory to calculate the splitting analytically. We find that the first-order energy correction comes from the effective RKKY-like interaction between the FMI and the impurity, while the second-order contribution originates from  triplet correlations present in an FMI/SC substrate.

The findings described above have direct consequences for the excitation spectrum causing a spin-splitting of the  lowest energy single-particle excitations, namely  the YSR excitations~\cite{yu,shiba,rusinov} in the strong coupling regime of the magnetic impurity. This splitting can be measured using tunneling probes on e.g. a quantum dot coupled to an FMI/SC nanowire or a magnetic adatom on the surface of FMI/SC heterostructure. 

The rest of article is organized as follows, in Sec.~\ref{sec:model}, we introduce the many-body Hamiltonian describing the system under study.  In Sec.~\ref{sec:NRG}, we describe the results of the NRG calculations of the ground state and the low energy excitations. In Section~\ref{sec:2site}, we introduce a minimal two-site model that qualitatively reproduces the key features of the results obtained using NRG. In Section~\ref{sec:PT}, we describe the results of perturbation theory applied to a two-site model and an extended superconductor in the limit of weak Kondo coupling. In Sec.~\ref{sec:spectral}, we describe the effect of the exchange field on other spectral features in the continuum part of the tunneling spectrum.  Unless otherwise stated, we will work in units where $\hbar = 1$.

\section{Model}
\label{sec:model}

\begin{figure}
    \centering
    \includegraphics[width=\columnwidth]{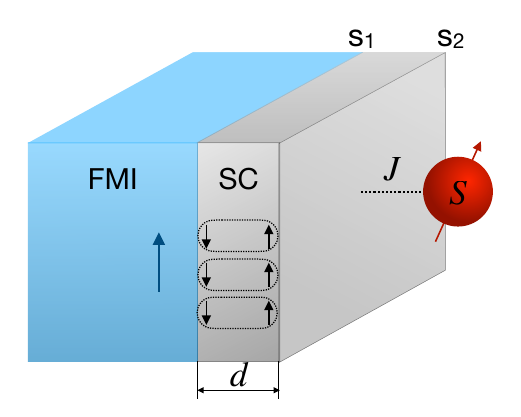}
    \caption{Schematic representation of an FMI/SC heterostructure coupled to an impurity spin $\bm S$ via a (Kondo) exchange coupling of amplitude $J$. The superconducting layer has a thickness of $d \ll \xi$, with $\xi$ the superconducting coherence length (but larger than the electron mean free path). FMI gives rise to the spin-splitting of the quasi-particle density of states in the SC and induces opposite spin polarization on the two superconducting interfaces $\text{S}_1$ and $\text{S}_2$, i.e. $\mathcal{P}(\text{S}_1) = -\mathcal{P}(\text{S}_2)$. This results in an effective RKKY-type interaction between the FMI and the impurity spin. The blue arrow indicates the orientation of the average magnetic moment in the FMI.}
    \label{fig:schematic}
\end{figure}
We consider a superconductor/ferromagnetic-insulator (FMI/SC) heterostructure coupled to a spin-$\tfrac{1}{2}$ magnetic impurity via an isotropic anti-ferromagnetic (Kondo) exchange interaction. We introduce the following Hamiltonian to describe the system:
\begin{align}
H &= H_{\text{0}} + H_{K} + H_{RKKY}, \label{eq:many-body-Hamiltonian} \\
H_{\text{0}} &= \sum_{\vec{k}\sigma}\epsilon_{\vec{k}} c_{\vec{k}\sigma}^\dag c_{\vec{k}\sigma } + 
 h\sum_{{\vec{k}},\sigma \sigma^{\prime}} \left[c_{\vec{k}\sigma}^{\dag} \left( 2 s^z_{\sigma\sigma^{\prime}}\right) c_{\vec{k} \sigma^{\prime}}   \right ] \notag\\&+\Delta \sum_{\vec{k}}\left[ c^\dag_{\vec{k}\downarrow}c^\dag_{-\vec{k}\uparrow}+c_{\vec{k}\downarrow}c_{-\vec{k}\uparrow}\right]\,\label{eq:h0},\\
H_{K} &= J \bm{S}\cdot{\bm s}_0\,, \\
H_{RKKY} &= J \rho_0h S^z\,.\label{eq:HRKKY}
\end{align}
Here, $c_{\vec{k} \sigma}$  $(c^{\dagger}_{\vec{k} \sigma})$ is the annihilation (creation) operator for the electrons in the superconductor. 

In Eq.~\eqref{eq:many-body-Hamiltonian}, $H_{\text{0}}$ describes the FMI/SC heterostructure in the absence of magnetic impurity. $\Delta$ is the strength of the pairing potential.
Proximity to an FMI gives rise to an exchange field $h$ in the SC that couples to the electron spin as a Zeeman field and therefore splits the Bogoliubov quasi-particle bands~\cite{Meservey:1991,meservey1994spin}. However, unlike the Zeeman coupling to a uniform magnetic field that equally splits all electronic bands, the effective exchange field stems from the magnetic proximity effect, and hence only splits the bands in an energy shell around the Fermi energy for which pairing correlations are important~\cite{tokuyasu1988proximity,hijano2021coexistence} (in conventional SCs, for excitation energy $\lesssim \omega_D$, $\omega_D$ being the Debye frequency). 

In most experiments where spin-splitting has been observed \cite{Moodera:1988,meservey1994spin,strambini2017revealing}, superconducting films, such as those made of aluminum, are used. These films have a thickness $d$ greater than the typical elastic mean-free path $\ell$ but smaller than the superconducting coherence length $\xi$. In this case, the exchange field at distances greater than $\ell$ from the FMI/SC interface can be considered homogeneous and proportional to $1/d$~\cite{tokuyasu1988proximity,heikkila2019thermal}.
From the mean-field perspective, the proximity-induced exchange field $h$ plays a role akin to the superconducting pairing potential. For simplicity, in Eq.~\eqref{eq:many-body-Hamiltonian} we choose $h$ to be oriented along the $z$-axis. 
 
$H_K$ describes an isotropic Kondo coupling between the impurity spin and the spin of the electrons in a superconductor, where $J$ is the exchange coupling, $\bm S$ the spin of the magnetic impurity and $\vec{s}_0$ the spin density of the superconductor at the position of the impurity, i.e. $\vec{s}_0 =(1/\Omega)\sum_{\vec{k} \vec{k}^{\prime}}c^\dagger_{\vec{k} \sigma} \vec{s}_{\sigma\sigma^{\prime}}  c_{\vec{k}^{\prime} \sigma^{\prime}}$, where $\vec{s} = \vec{\sigma}/2$ are (half) the spin Pauli matrices and $\Omega$ the system volume. The energy scales that determine the existence of a magnetic moment at the impurity and its interaction with the host electrons are the intra-orbital Coulomb energy $U$, and the orbital energy $\epsilon_d$, which are typically much larger than $\omega_D$. Between $U$ and $\omega_D$, pairing fluctuations are negligible and the electronic states are not spin-split by the proximity to the FMI. Therefore, at the scale where the pairing and exchange potentials set in, the Kondo coupling can be considered spin-isotropic. 

However, proximity to the FMI modifies the interaction of the impurity spin with the superconductor by giving rise to an effective RKKY-type of interaction described by $H_{RKKY}$ in Eq.~\eqref{eq:many-body-Hamiltonian}. In the presence of the FMI, the superconductor is locally polarized at the interface $\text{S}_1$ (see Fig.~\ref{fig:schematic}). Since the impurity is located at the interface $\text{S}_2$, at a distance $d$ ($d\ll \xi$ but $d\gg\ell$) to the FMI, the surface $\text{S}_2$ exhibits opposite spin polarization due to pairing correlations (see Fig.~\ref{fig:schematic}). Therefore, an RKKY exchange field emerges at the interface $\text{S}_2$, which is modeled by $H_{RKKY} = J\langle \vec{s}\rangle_{\text{S}_2}\cdot {\bm S} $ where $\langle \vec{s}\rangle_{\text{S}_2}$ is the polarization at $\text{S}_2$ and it is estimated to leading order as the Pauli susceptibility\cite{kochelaev1979spatial,izyumov2002competition,bergeret2004spin}, i.e. $\langle \vec{s}\rangle_{\text{S}_2}=-\langle \vec{s}\rangle_{\text{S}_1}=\chi_S \bm{h} \simeq \rho_0 \bm{h} $ where $\bm{h} = h \bm{e_z}$, $\chi_S$ is the spin susceptibility and $\rho_0$ is the density of state at the Fermi surface.

Taking into account all these considerations and neglecting any weak scattering potential that breaks particle-hole symmetry, the total Hamiltonian describing the system is given by Eq.~\eqref{eq:many-body-Hamiltonian}. We stress that it is implicitly understood that this model is an effective description of a magnetic impurity at the surface of the FMI/SC system at energy scales $\approx \omega_D$ or lower.
The eigenstates of this model Hamiltonian can be labeled by the $z$-component of the total spin, i.e. $S^z_T = S^z + \sum_{\vec{k},\sigma\sigma^{\prime}} c^{\dag}_{\vec{k} \sigma} s^{z}_{\sigma\sigma^{\prime}} c_{\vec{k}^{\prime} \sigma^{\prime}} $ and the fermion parity operator. The latter is defined as
$P = \prod_{\vec{k}, \sigma} (-1)^{n_{\vec{k}\sigma}} \,,$ where $n_{\vec{k}\sigma} = c^\dagger_{\vec{k}\sigma}c^{ }_{\vec{k}\sigma}$.

In the absence of the FMI, i.e. for $h=0$ and $H_{RKKY}  =0$, this model has been extensively studied in the past and its ground-state and low-lying excitation spectrum are fairly well understood~\cite{yu,shiba,rusinov,Balatsky_2006}: As a function of the exchange coupling $J$, the many-body ground state undergoes a (level-crossing) phase transition from a doublet  at $J<J_{\text{cr}}$ (weak coupling regime) to a Kondo singlet ground-state at $J>J_{\text{cr}}$ (strong coupling regime). At $J<J_{\text{cr}}$ the impurity spin is weakly coupled to the electrons in a superconductor and the ground state of the system is two-fold degenerate \textit{doublet} state that is well approximated as a product state of the BCS ground state and the impurity-spin states, i.e. $|\text{BCS} \rangle |\pm \!1/2\rangle$. Thus, in the weak coupling limit, the ground-state doublet has even parity and its total spin is equal to the impurity spin, i.e. $\langle S^z_T \rangle  = \langle S^z\rangle = \pm 1/2$. As the exchange coupling $J$ increases, the spin of the impurity couples more strongly to the local spin fluctuations in the superconductor. Eventually, for $J>J_{\text{cr}}$ the ground-state energy is lowered by forming a collective singlet bound state 
(i.e. $\langle S^z_T \rangle = 0$) consisting of the impurity spin and a polarization cloud of quasi-particle excitations from the superconductor. In the strong coupling limit the ground state has odd fermion parity and it is a singlet~\cite{JPSJ.67.1332_BCS_trans,JPSJ.67.1332_BCS_trans2,JPSJ.69.1812_BCS_trans3,Balatsky_2006}. 

When the single-particle excitation spectrum is probed by a tunneling probe, the spectrum displays a pair of narrow peaks at subgap energies symmetrically around zero bias (i.e. $E = 0$). They correspond to the excitations known as the Yu-Shiba-Rusinov (YSR) states~\cite{yu,shiba,rusinov} and can be excited by a single tunneling electron for $E>0$ (or hole for $E<0$). Tunneling of electrons or holes couples states of different fermion parity. Indeed, YSR excitations have previously been described as eigenstates of the Bogoliubov-de Gennes equations obtained from the quadratic Hamiltonian that results after replacing the impurity spin operator with a classical vector. However, this classical description neglects quantum fluctuations of the impurity spin~\cite{JPSJ.67.1332_BCS_trans,JPSJ.67.1332_BCS_trans,Lobos:2019} and important many-body effects related to e.g. parity of the ground state~\cite{JPSJ.67.1332_BCS_trans,JPSJ.67.1332_BCS_trans2,von-oppen:2021,trivini2022pair} or the spin of the YSR excitations~\cite{skurativska:2023}.

Placing a thin superconducting layer in proximity to the FMI brings about new features that are not encountered in conventional SCs. The exchange field $h$ induced in the superconductor via the magnetic proximity effect generates superconducting correlations in the triplet state, which are `odd' in the Matsubara frequency and are not present when $h = 0$ \cite{bergeret2005odd,bergeret:2023}.
Here, we investigate how triplet correlations affect the properties of the magnetic impurity.

Below, we argue that the existence of the triplet correlations have important consequences for the low-lying spectrum of the system described by Eq.~\eqref{eq:many-body-Hamiltonian}. Indeed, even if the stability of the (spin-split) superconductor requires $|h| < \Delta$, the exchange field cannot be treated as a weak perturbation. It substantially alters the quasi-particle spectrum below the energy scale for which superconducting correlations are important ($\approx \omega_D$), and we need to treat its effects using a non-perturbative method. This is accomplished by using the numerical renormalization group (NRG) and the results are described in the following section.

\section{NRG results}
\label{sec:NRG}
\begin{figure}[h]
    \centering
    \includegraphics[width=\columnwidth]{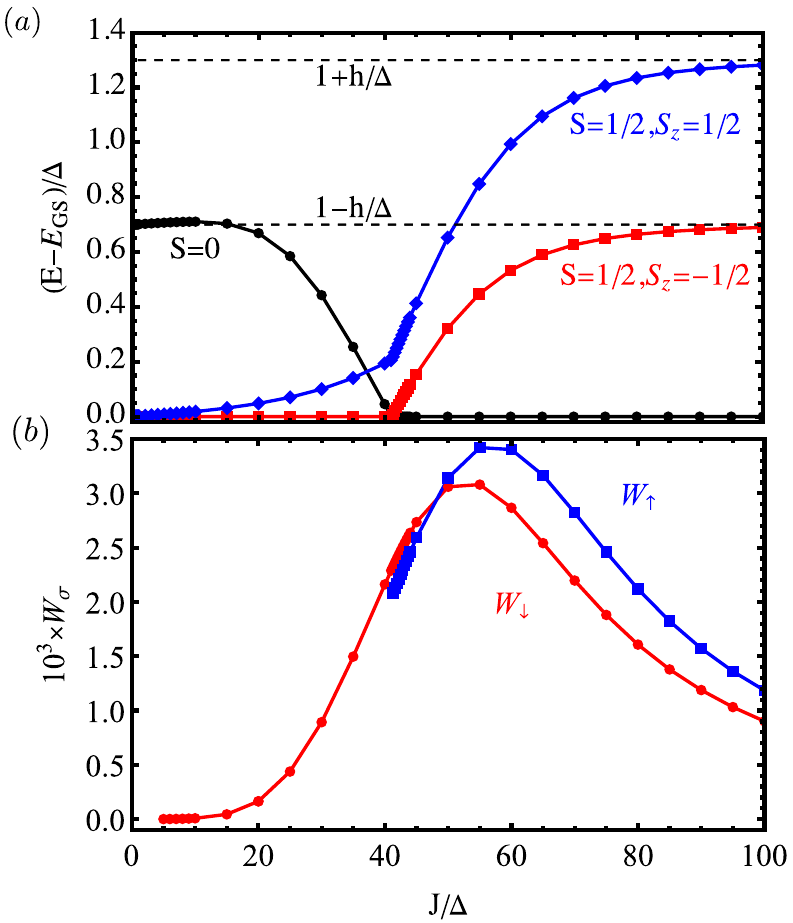}
    \caption{(a) Energies of the low-lying states of the Hamiltonian in Eq.~\eqref{eq:many-body-Hamiltonian} as a function of exchange coupling $J$ in the presence of exchange field, $h= 0.3\Delta$. The black line corresponds to the singlet state, and the blue and red lines are the doublet states split by the exchange field. (b) Spectral weights of the YSR excitations as a function of exchange coupling. The blue and red lines are the spin-resolved spectral weights of the spin-up and -down components at the exchange field, $h=0.3\Delta$.  
    }
    \label{fig1:nrg-results}
\end{figure}

In this section, we solve the model in Eq.~\eqref{eq:many-body-Hamiltonian} using the numerical renormalization group (NRG) method~\cite{RevModPhys.47.773_NRG,JPSJ.67.1332_BCS_trans,JPSJ.67.1332_BCS_trans2,JPSJ.69.1812_BCS_trans3,RevModPhys.80.395_NRG}.  

Tunneling probes like the scanning tunneling microscope (STM) can access the single-particle spectrum of the system. The tunneling current is proportional to the convolution of the spectral function of the system with that of the tunneling probe (e.g. the tip of the STM). The definition and calculation using NRG of the spectral function, $A_{\sigma}(\omega)$, are described below in  Sec.~\ref{sec:spectral} and in App.~\ref{app:nrg}. Here, we shall simply outline the key results.

\begin{figure}[t]
\includegraphics[width=\columnwidth]{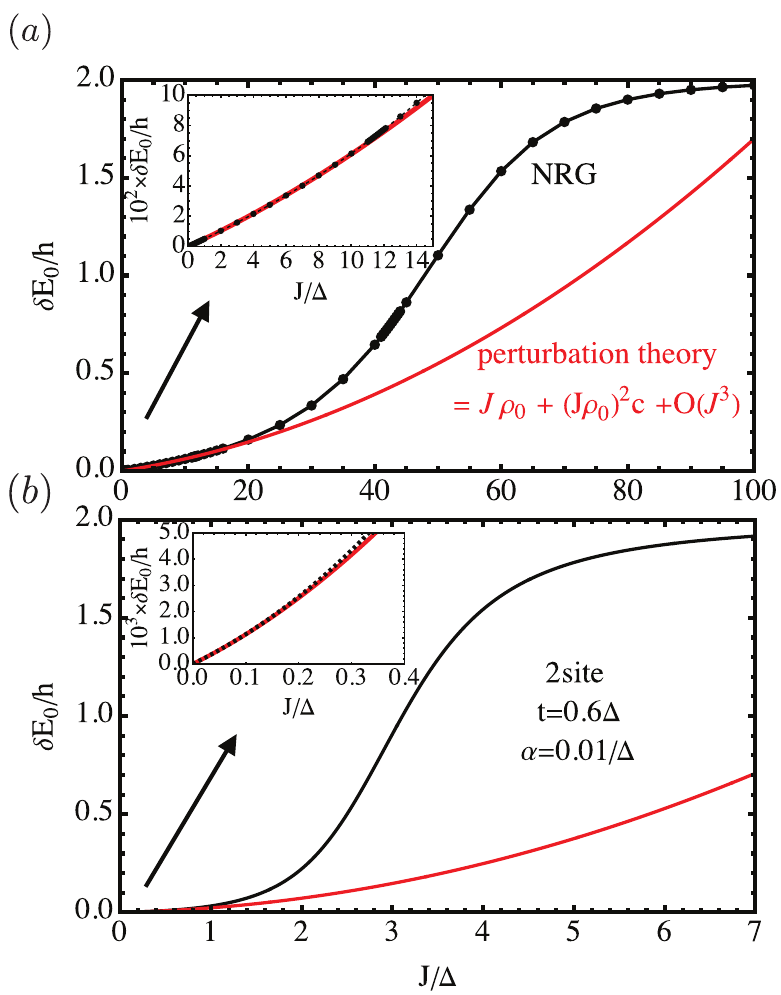}
\caption{Ratio of the energy splitting of the two low-lying even parity states $\delta E_0$ to the strength exchange field $h$ as a function of Kondo  coupling $J$ obtained via NRG (a) and the minimal two-site model of Eq.~\eqref{eq:two-site-model} (b). 
In the weak coupling regime the splitting is well reproduced by a linear + quadratic law derived using perturbation theory (red curves). For the NRG, the non-universal constant $c\simeq4.43$ is obtained by curve fitting. Parameters used for the NRG calculations: $D=1$, $\Delta=0.01 D$, $h=0.3\Delta$. Parameters used for the two-site model: $\Delta=0.2$, $t=0.6\Delta$, $h=0.3\Delta$.}\label{fig2:splitting-NRG-two-site}
\end{figure}

Figure~\ref{fig1:nrg-results}(a) shows the evolution with the Kondo coupling $J$ of the singlet (in black) and doublet (in blue and red) excitation energies taking the strength of the pairing potential in Eq.~\eqref{eq:many-body-Hamiltonian} to be $\Delta = 10^{-2} D$, where $D \simeq \omega_D$ is the energy cutoff or bandwidth of the model. In Fig.~\ref{fig1:nrg-results} $E_{GS}$ refers to the ground state energy. Note, the ground state is different in the weak and strong coupling regimes.  
 The level crossing quantum phase transition from the weak coupling regime to the strong coupling regime takes place at $J/\Delta\simeq 40$. In Figure~\ref{fig2:splitting-NRG-two-site}(a) we show the value of splitting computed from NRG. The value of splitting saturates to $2h$ at strong coupling. At weak coupling, the splitting is linear in both $J$ and $h$ (see insert of Fig.~\ref{fig2:splitting-NRG-two-site}). This behavior is well reproduced by the leading order perturbation theory, as discussed in Section~\ref{sec:PT}.
 {Within perturbation theory, the splitting is given by  $\delta E_0/h=J\rho_0 + (J\rho_0)^2 c +O(J^3)$ where $c$ depends logarithmically on the bandwith $D$ and the gap $\Delta$. The first-order term stems solely from the RKKY interaction and the second-order term is due to the triplet correlations built in the spin-split superconductor. Qualitatively, all of these features at strong and weak coupling are also captured by the minimal two-site model introduced in the next section. }

Fig.~\ref{fig1:nrg-results}(b) shows the evolution of the spectral weight of the YSR peaks in the spectral function. Notice, in the weak coupling regime for $h\neq 0$ the absolute ground state has total spin  $S^z_T = -\frac{1}{2}$ (see figure~\ref{fig1:nrg-results}(a)). Since a tunneling electron (hole) can only couple states of the opposite fermion parity, in the weak coupling regime \emph{at zero temperature}, the YSR excitation is only due to the transition from the spin-down even parity ground state to the singlet state. On the other hand, in the strong coupling regime, the ground state is the odd parity singlet,  and therefore the YSR  corresponds to excitations to the two low-lying states of even parity states, which are split by the exchange field $h$. As a consequence,  whilst the spin-down spectral weight (red line) is continuous across the transition,  the spin-up spectral weight (blue line)  is non-zero only in the strong coupling regime.
In all cases, it is also worth pointing out, as already noticed in Ref.~\cite{skurativska:2023}, that the YSR states exhibit a robust spin polarization.  


\section{Two-site model} \label{sec:2site}

NRG provides a quantitatively accurate description of the low-lying states and single-particle spectrum of the model in Eq.~\eqref{eq:many-body-Hamiltonian}. However, it does not shed much light on the physical origin of various spectral features. In this section, we show that some understanding of the latter can be obtained by studying a minimal model consisting of a magnetic impurity coupled to two superconducting sites. Previously, Ref.~\cite{skurativska:2023} modeled a spin-$\tfrac{1}{2}$ magnetic impurity coupled to the FMI/SC heterostructure using a single-site model. In this model the spin-split superconductor is described by a single fermion site coupled to a spin-$\tfrac{1}{2}$ impurity. Although this model correctly predicts that the YSR states are spin-polarized, it cannot describe the variation of the spin-splitting with $J$ that is observed in the NRG calculations (see Fig.~\ref{fig1:nrg-results}(a)). In the following, we show that a minimal model capturing this and other effects obtained using NRG requires necessarily two sites to describe the spin-split superconductor. In addition, we demonstrate that the splitting is the consequence of the magnetic RKKY interaction and the existence of spin-triplet correlations in the spin-split superconductor~\cite{bergeret:2023}. 

\begin{figure*}[t]
\includegraphics[width=\textwidth]{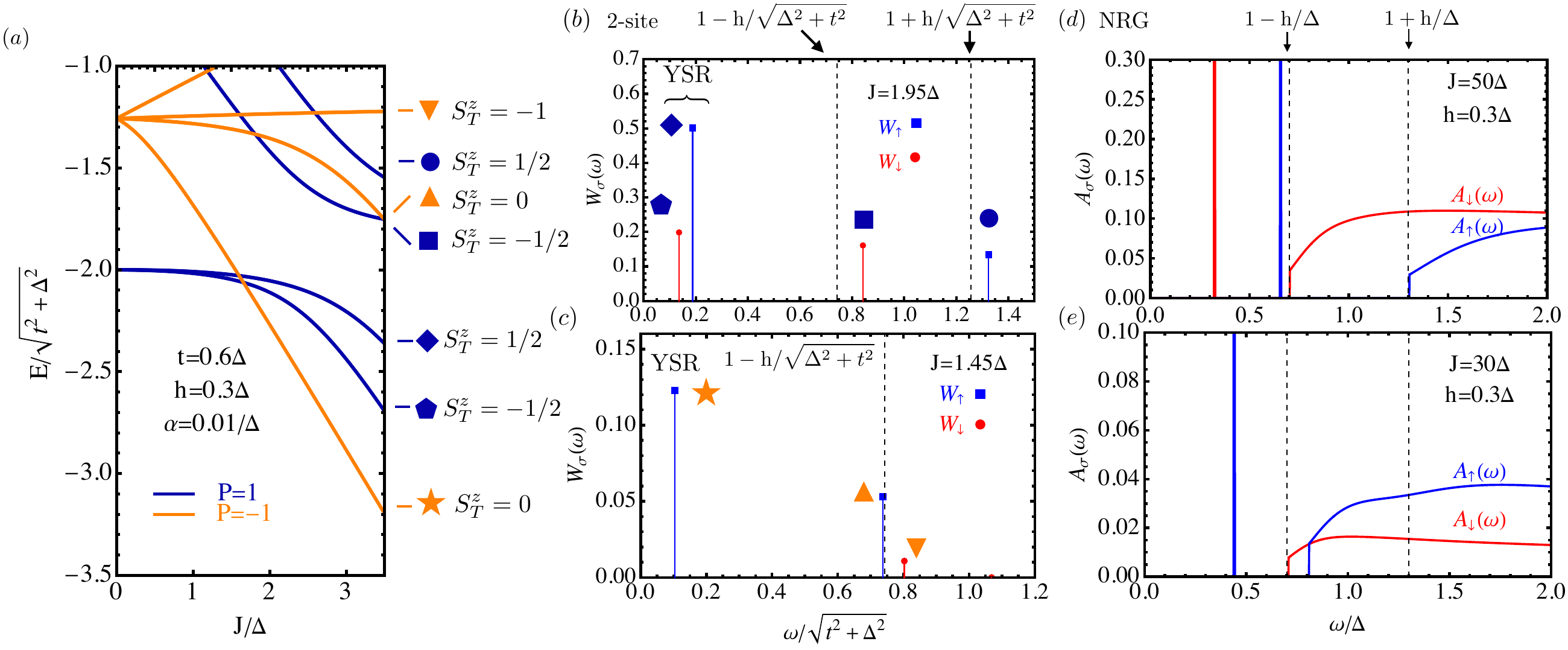}
\caption{(a) Low-energy spectrum of the two-site model (cf. Eq.~\ref{eq:two-site-model}) labeled by the fermion parity $P = \pm 1$ and $z$-component of the total spin $S^z_T$. 
Panels (b, c) are the spectral weights in the strong and weak coupling regimes, respectively of the discrete excitation peaks, i.e. $W_{\sigma}(\omega) = \lim_{\epsilon\to 0}\int^{\omega+\epsilon}_{\omega-\epsilon} d\omega^{\prime} \: A_{\sigma}(\omega^{\prime})$, where $A_{\sigma}(\omega)$ is the spin-resolved spectral
function of the two-site model (see main text for details). 
Panels  (d, e) are plots of $A_{\sigma}(\omega)$ for the full model in the strong and weak coupling regimes, respectively. The dashed vertical lines indicate the positions of the threshold for continuum excitations in the classical model described in Ref.~\cite{skurativska:2023}. Red and blue correspond to spin-up and spin-down components of $A_{\sigma = \{ \uparrow,\downarrow \} }(\omega)$, respectively.}  \label{fig3:NRG-vs-2site}
\end{figure*}

In the strong coupling regime, the splitting of the lowest-lying even-parity states approaches twice the exchange energy, $2h$ (see Fig.~\ref{fig2:splitting-NRG-two-site}(a)). It may appear as counter-intuitive that strongly coupled impurity is sensitive to the exchange field. However, as shown by the NRG method and confirmed by the two-site model introduced below, in the strong coupling regime the system exhibits a pair of spin-split single-particle excitations that correspond to the YSR peaks of the spectral function (cf. Fig.~\ref{fig3:NRG-vs-2site}(c-e)). 
%
\subsection{Description of the model}
Without further ado, let us introduce a two-site model that describes the spin-split superconductor as two sites with the same $s$-wave pairing ($\propto \Delta$) and exchange ($\propto h$) potentials. Tunneling of the electrons between the two sites with amplitude $t$ is also allowed. The impurity spin $\bm S$ is coupled via Kondo  coupling $J$ to the spin of the electrons on one of the two superconducting sites described by ${\bm s}_0$. The Hamiltonian reads
\begin{align}
H &= H_0 + H_K+H_{RKKY} \label{eq:two-site-model}\,,\\
H_0 &= \sum_{i = 0, 1} \left[ \Delta ( c^\dagger_{i \uparrow} c^\dagger_{i \downarrow} + \text{H.c.} ) +  h \, c^{\dag}_{i \sigma} \left( 2 s^z_{\sigma\sigma^{\prime}} \right) c_{i \sigma^{\prime}}\right] \notag\\ & \qquad -  t (c^\dagger_{0\sigma} c_{1\sigma} + \text{H.c.} ) \notag\,,\\
H_K &= \frac{J}{2}(S^+s^-_0 +\text{H.c.}) + J S^z s^z_0  \notag\,,\\
H_{RKKY} &= J h \alpha S^z \notag\,.
\end{align}
In the above expression, we have explicitly written the spin-flip and Ising parts of the exchange interaction by introducing $S^{\pm} = S^x \pm i S^y$, $s^{\pm}_0 = s^{x}_0\pm i s^{y}_0$, etc. Note that in the two-site model, the spectrum is discrete, and therefore there is no Pauli paramagnetism. Thus, we have replaced the density of states at the Fermi energy, $\rho_0$, by a phenomenological constant which parametrizes the strength of the RKKY interaction between the impurity and the FMI. 

The Hilbert space of this model has a relatively small dimension ($=32$) and thus full exact diagonalization of the Hamiltonian~\eqref{eq:two-site-model} is possible. Moreover, the numerical
calculations are aided by the existence of a few conserved quantities. 
The fermion parity operator takes the form $P=(-1)^{\sum_{i\sigma}{n_{i\sigma}}},$ where $n_{i\sigma} = c^\dagger_{i\sigma} c_{i\sigma}$ ($i=0,1$). The system is also invariant under a global spin rotation around the $z$-axis and therefore, along with $P$, $S^z_T = S^z + \sum_{i=0,1} c^{\dag}_{i \sigma} s^z_{\sigma\sigma^{\prime}} c_{i\sigma^{\prime}}$ is also conserved. Thus, the eigenstates of the Hamiltonian in Eq.~\eqref{eq:two-site-model} can be labeled by their fermion parity: even ($P=+1$) or odd ($P=-1$), and the eigenvalue of $S^z_T$. 
\subsection{Exact diagonalization results}
We demonstrate that the two-site model qualitatively reproduces the key spectral properties found in NRG. We obtain the spectrum of the two-site model by diagonalizing the Hamiltonian in Eq.~\eqref{eq:two-site-model}. Fig.~\ref{fig3:NRG-vs-2site}(a) shows the low-energy spectrum of the two-site model as a function of the exchange coupling $J$. When $J\!=\!0$ the system is in the parity-even two-fold (doublet) degenerate ground state with eigenvalue of $S^z_T$ ($S^z$) equal to $\pm 1/2$ and the ground-state energy is $E_0=-2\sqrt{\Delta^2+t^2}$. In the basis of bonding and anti-bonding creation (annihilation) operators $c^{(\dagger)}_\pm = (c^{(\dagger)}_0 \pm c^{(\dagger)}_1)/\sqrt{2}$ the doublet is $| \text{GS}_m \rangle = |\text{BCS}\rangle_+|\text{BCS}\rangle_-|m=\pm \frac{1}{2}\rangle$, where $|\text{BCS}\rangle_{\pm} = (u_{\pm}|0\rangle_{\pm} +v_{\pm}|2\rangle_{\pm})/\sqrt{2}$, the expressions of $u_{\pm}$ and $v_{\pm}$ in terms of $\Delta,t,h$ can be found in Appendix~\ref{app:B}.   

The model has two different ground states depending on the strength of the exchange coupling $J$. In the weak coupling regime ($J<J_{\text{cr}}$) the impurity spin is weakly entangled with the superconducting sites and remains unscreened. The ground state is adiabatically connected with one of the states of the $J = 0$ doublet, that is a product of even-parity states. As $J$ increases the Kondo singlet which roughly corresponds to an entangled state of the impurity spin and a quasiparticle localized in the two sites is gradually pushed down in energy (see Fig.~\ref{fig3:NRG-vs-2site}(a)). Eventually, at $J = J_{\text{cr}}\simeq 1.5 \Delta$, the system undergoes a level-crossing phase transition changing the fermion parity of the ground state from even to odd, such that the singlet state eventually becomes the ground state in the strong coupling limit ($J>J_{\text{cr}}$). 
 
Notice that for $h = 0$, time-reversal symmetry is restored and the (Kramers) degeneracy of the ground state in the weak coupling limit is guaranteed. In this regime, provided there is no bias in the preparation of the system, the ground state is described by a mixed state with equal (classical) probability for the two states of the doublet~\cite{skurativska:2023}. However, setting $h \neq 0$ breaks time-reversal symmetry and lifts the degeneracy of the even-parity ground state doublet. Thus, as found in the NRG results for the full model~\eqref{eq:many-body-Hamiltonian}, in the two-site model for $J < J_{\text{cr}}$ we also observe that the degeneracy of the $h=0$ doublet is lifted and a finite splitting between the two lowest-lying even parity states (denoted $\delta E_{0}$ from here on) appears. Notice that $\delta E_0\neq 0$ even for small $J$ and it increases with $J$ up to a maximum $\Delta E_{0} = 2h$, that is, twice the Zeeman energy caused by the exchange field. Therefore, the behavior of $\delta E_0$ qualitatively reproduces the NRG results (compare the two panels of Fig.~\ref{fig2:splitting-NRG-two-site}).  Furthermore, in the two-site model as shown below analytically, the splitting is linear in $J$ at vanishing hoping amplitude (i.e. for $t=0$). In this limit, the two-site model reduces to a single-site model studied in Ref.~\cite{skurativska:2023} in an RKKY-like (mangetic) field $\propto J$. This model 
fails to reproduce the spectral features observed in NRG. 

 The two-site model offers a rather simple explanation for the saturation of the splitting $\delta E_0$ to twice the Zeeman energy at large Kondo coupling, i.e. $\delta E_0(J\to +\infty) \to 2h$  (cf. Fig.~\ref{fig2:splitting-NRG-two-site}). This saturation
is observed both in NRG and in the two-site model and can be understood as follows: For large $J$, the impurity captures a quasi-particle and localizes it in the first site of the chain (i.e. the $i=0$ site). The captured quasi-particle and the impurity form a tightly bound Kondo singlet with binding energy $\sim -\tfrac{3}{4} J$ for large $J\gg t,\Delta,h$. In this limit, the RKKY interaction acting on the impurity spin alone has zero expectation value, and hopping from the site $i=1$ into the site $i=0$ and vice versa is also suppressed. Thus, the single-particle excitations of the system must ``live'' on the site(s) with $i > 0$. If we neglect the hopping amplitude $t$ in this large $J$ limit, the lowest quasi-particle excitations at energy $\Delta$ are spin-split due to the exchange field by an amount equal to $2h$. Notice that, since the (Kondo) singlet ground state has odd fermion parity, the lowest-energy quasi-particle excitations over this ground state have even parity, and therefore $2h$ is the smallest spin splitting of even parity states for $J\to +\infty$.

\section{Perturbation theory}\label{sec:PT}

In what follows, we use perturbation theory to analytically obtain the splitting 
$\delta E_0$ in the weak coupling regime where $J \ll \Delta, h, t$. In this regime, there are two contributions responsible for the splitting of the lowest-lying even-parity states: the effective RKKY interaction induced by the FMI, which is of $O(J)$ and the different kind of triplet correlations built into the spin-split superconductor, which appears at $O(J^2)$. 

\begin{align}
\delta E_0 = \delta E_0^{(1)}+ \delta E_0^{(2)}\,,
\end{align}
where the term linear in both $J$ and $h$ is given by
\begin{align}
\label{eq:pt-1st-order}
    \delta E_0^{(1)}&= \langle GS_{m=+\tfrac{1}{2}}|H_{RKKY}| GS_{m=+\tfrac{1}{2}}\rangle \\
   &\quad  - \langle GS_{m=-\tfrac{1}{2}}|H_{RKKY}| GS_{m=-\tfrac{1}{2}}\rangle\,,
\end{align}
For the calculation of the second-order contribution, the starting point is the following expression of the energy splitting of the lowest-lying even-parity  states whose derivation is discussed in Appendix~\ref{app:split}:
\begin{equation}
\delta E^{(2)}_0 =  \frac{J^2}{4} \left(\chi^{+-} - \chi^{-+} \right), \label{eq:spliteq}
\end{equation}
where
\begin{align}
\chi^{ab} &= -\int^{+\infty}_0 d\tau\,  C^{ab}(\tau)\\
&= \sum_{E} \frac{\langle \text{BCS} | s^a | E\rangle \langle E | s^b|  \text{BCS}\rangle }{E - E_0}\label{eq:sume}, \\
\mathcal{C}^{ab}(\tau) &= -\langle \text{BCS} |\mathcal{T} \left[ s^{a}(\tau) s^b(0)\right] | \text{BCS} \rangle\,.
\end{align}
Eq.~\eqref{eq:sume} is most useful when dealing with the finite-size system described 
by Eq.~\eqref{eq:two-site-model}. The expression in terms of the time-ordered spin correlation 
functions $\mathcal{C}^{ab}(\tau)$ will be useful for carrying out the perturbative calculation
in the case of an extended (infinite) superconductor. In Eq.~\eqref{eq:sume}, $|\text{BCS}\rangle$ and $|E\rangle$ are, respectively, the ground state excited states of the superconductor Hamiltonian $H_0$ (cf. Eqs.~\eqref{eq:many-body-Hamiltonian} and ~\eqref{eq:two-site-model}). 

\subsection{Two-site model} \label{sec:two-sitepert}

Let us first consider the two-site model.
The first-order in $J$ correction to ground-state energy comes from RKKY term $H_{RKKY}$ in the Hamiltonian \eqref{eq:two-site-model} and it reads
\begin{equation}
\label{eq:two-site-1st-order}
\delta E^{(1)}_0 = \pm \frac{1}{2}Jh \alpha\,. 
\end{equation}
In order to compute $\chi^{+-}$ and $\chi^{-+}$, we use
\begin{equation}
s^{\pm}_0 |\text{BCS}\rangle =  \mp \alpha |t_{\pm 1}\rangle\,,
\end{equation}
where $\alpha$ is a function of $t,\Delta, h$ (for the details see Appendix \ref{app:B}),
$|\text{BCS}\rangle = |\text{BCS}\rangle_{+} |\text{BCS}\rangle_{-},$ and we have introduced the spin-triplet quasi-particle states:
\begin{align}
    |t_{+1}\rangle &= |\!\uparrow\uparrow\rangle =\gamma^{\dag}_{
+,\uparrow}\gamma^{\dag}_{-,\uparrow}|\text{BCS}\rangle\,, \\
  |t_{0}\rangle &=   \frac{(|\!\uparrow \downarrow\rangle + |\!\downarrow \uparrow \rangle)}{\sqrt{2}},\notag\\
  &= \frac{1}{\sqrt{2}}\left[ \gamma^{\dag}_{+,\uparrow} \gamma^{\dag}_{-\downarrow} + 
   \gamma^{\dag}_{-,\uparrow} \gamma^{\dag}_{+,\downarrow}\right] |\text{BCS}\rangle, \\
  |t_{-1}\rangle &= |\!\downarrow\downarrow \rangle = 
  \gamma^{\dag}_{+,\downarrow}\gamma^{\dag}_{-,\downarrow}|\text{BCS}\rangle\,,
  \label{eq:spin-triplets}
  \end{align}    
where $\gamma_{\pm,\sigma}$ ($\gamma^{\dag}_{\pm,\sigma}$) destroys (creates) a quasi-particle  in the bonding ($-$)
and anti-bonding ($+$) orbitals with spin $\sigma$. The triplet states have eigenenergies $E = 
\{E_{t_{+1}},E_{t_{0}},E_{t_{-1}}\} =  \{2h, 0, -2h \}$ 
respectively. The ground state energy is $E_0 = -2\sqrt{\Delta^2 + t^2}$.

To the lowest order in perturbation theory   
the doublet $|\text{GS}_{m}\rangle = |\text{BCS}\rangle|m=\pm\tfrac{1}{2}\rangle$
is coupled to spin-triplet quasi-particle excitations via the Kondo 
exchange $H_K$. Since the energy of $|t_{\pm 1}\rangle$ is split by
the exchange field, this results in the following splitting of the lowest-energy
even parity states:
\begin{align}
\label{eq:splitting-two-site}
    \delta E^{(2)}_{0}&=\frac{J^2 t^2h}{16(t^2+\Delta^2)(\Delta^2-h^2+t^2)},\\
   & \simeq \frac{J^2 t^2h}{16(t^2+\Delta^2)^2} + O(h^3)\,,
\end{align}
to the leading order in $J$ and $h$. Notice, the splitting in Eq.~\eqref{eq:splitting-two-site} vanishes either in the absence of exchange field $h=0$ or for $t=0$, which is the limit of a single-site model, in agreement with the previous findings in Ref.~\cite{skurativska:2023}.

To summarize, in the weak coupling limit, the two-site model shows the splitting of the two lowest-lying even-parity states. In this model, the contribution of $O(J^2)$ to the splitting originates from the coupling of the doublet states at $J=0$ to the spin-split triplet states containing two quasi-particles. This result also clarifies why the single-site model of Ref.~\cite{skurativska:2023} does not describe this second-order contribution to the splitting $\delta E^{(2)}_0$. Indeed, the minimum number of independent orbitals required to construct a spin-triplet with two quasi-particles is two (the bonding and anti-bonding orbitals $\pm$ for the two-site model).

\begin{figure}[H]
\includegraphics[width=\columnwidth]{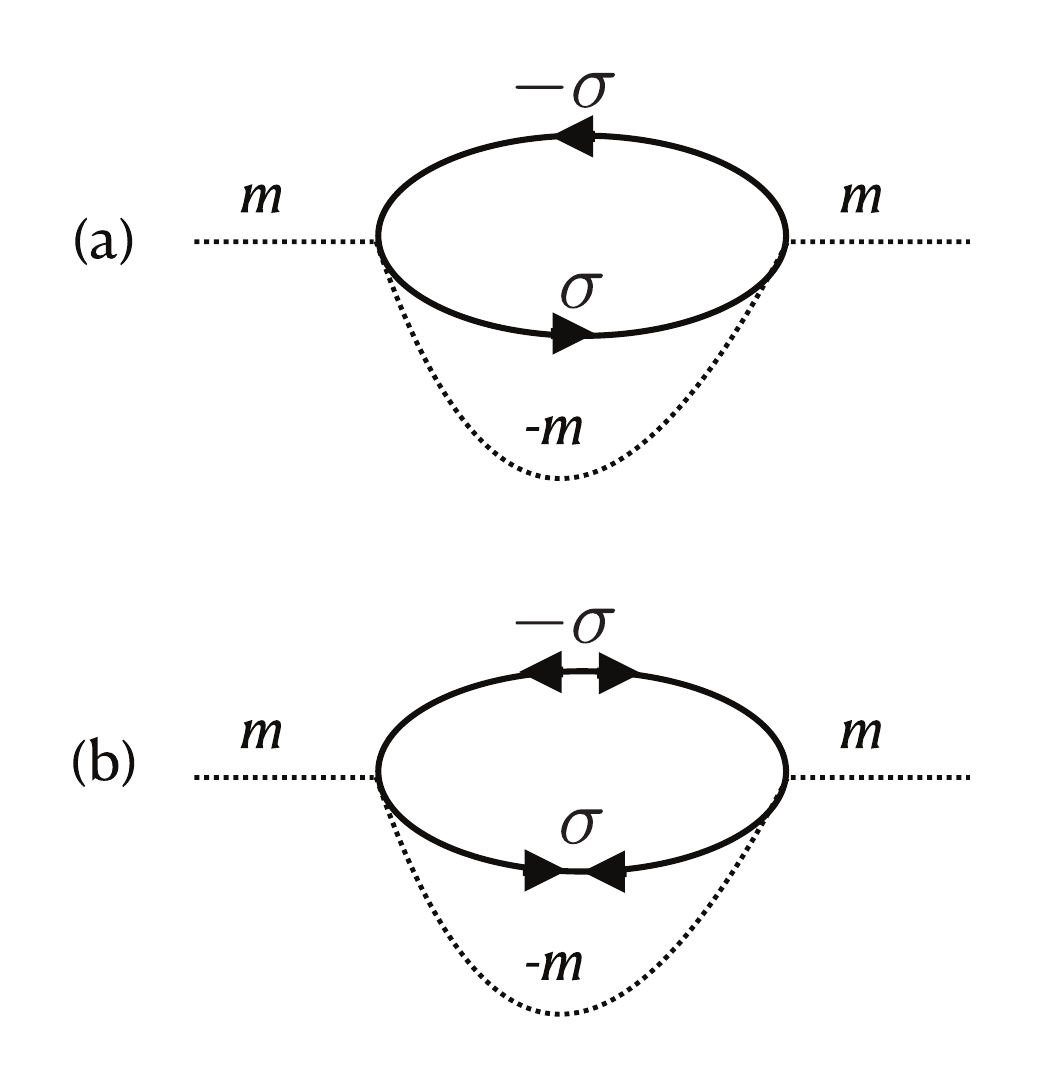}
\caption{Normal (a) and anomalous (b) Feynman diagrams contributing in second-order perturbation theory to the energy splitting of the two lowest-lying eigenstates with even fermion parity. The energy splitting is the difference between the $m=+\tfrac{1}{2}$ and $m=-\tfrac{1}{2}$ diagrams. The dashed lines describe the impurity state and the continuous lines correspond to the superconductor. The normal contribution (a) is logarithmically divergent. See Appendix~\ref{app:split} and \ref{app:logs} for details.}\label{fig:feynman}
\end{figure}

\subsection{Extended superconductor}

In the case of the extended superconductor, the first-order contribution to the splitting is determined by the expectation value of $H_{RKKY}$ according to Eq.~\eqref{eq:pt-1st-order}. This yields
\begin{equation}
\delta E^{(1)}_0 = J h \rho_0.  
\end{equation}
Concerning the contribution of the order $O(J^2)$, starting from Eq.~\eqref{eq:spliteq} the second order contribution can be written in terms of singlet and triplet components of the local single-particle Green's function $\mathcal{G}^{0}(i \omega)$ as follows:
\begin{align}
\delta E^{(2)}_0 = \frac{J^2}{4} \int \frac{d\omega d\omega^{\prime}}{2\pi}  \frac{\mathrm{Tr}\left[ g_s(i\omega) g_t(i\omega-i\omega^{\prime})\right]}{i\omega},\label{eq:e0}
\end{align}
where the $2\times 2$ Nambu matrices $g_s(i\omega)$ and $g_t(i\omega)$ are the singlet and triplet components of $\mathcal{G}^0(i\omega)$ (see Eqs.~\eqref{eq:g0st} and \eqref{eq:gpm} in Appendix~\ref{app:tripletcorr} for explicit expressions). Since the triplet component $g_t(i\omega)\to 0$ for $h\to 0$, the above result clearly shows that $\delta E^{(2)}_0$ vanishes in the absence of the triplet correlations. Interestingly, in the opposite strong coupling regime where the above perturbative treatment does not apply, the triplet correlations encoded in $g_t$ are also responsible for the splitting of the quasi-particle bands which results in $\delta E_0 \to 2h$. This intuition is confirmed by the NRG analysis for the extended model and the exact diagonalization results for the two-site model shown in Fig.~\ref{fig2:splitting-NRG-two-site}: The energies of the two lowest-lying even-parity states evolve smoothly from the weak to the strong coupling regime. This is also important from the experimental point of view, because in the excitation spectrum, the splitting can only be observed in the strong coupling regime, see center and right panels of Fig.~\ref{fig3:NRG-vs-2site}.

\begin{figure}[t]
\includegraphics[width=\columnwidth]{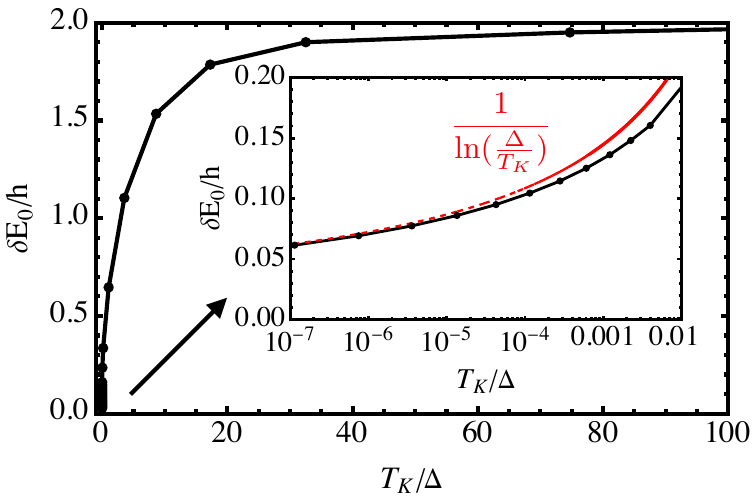}
\caption{Ratio of the energy splitting  of the two lowest even-parity states $\delta E_0$ to the exchange field of the ferromagnetic insulator $h$  vs. the ratio of the Kondo temperature $T_K$ to the superconducting gap $\Delta$. The Kondo temperature is extracted from the half-width at half-maximum~\cite{PhysRevLett.85.1504_Kondo} of the  Kondo peak computed from NRG for a spin-$\tfrac{1}{2}$ impurity with Kondo coupling $J$ to a normal metal. For $T_K/\Delta \ll 1$, the splitting behaves as $1/\ln(\Delta/T_K)$ and saturates to 2 for $T_K/\Delta \gg 1$. } \label{fig:scaling}
\end{figure}

The expression in Eq.~\eqref{eq:e0} can be evaluated explicitly and in the wide band limit, with logarithmic accuracy, it takes the form:
\begin{equation}
\delta E^{(2)}_0 \simeq (J\rho_0)^2  h  \ln \left( \frac{D}{\Delta} \right),\label{eq:2split}
\end{equation}
where $D$ ($\sim \omega_D$ in conventional superconductors) is the bandwidth of the superconductor described by $H_0$ term in Eq.~\eqref{eq:many-body-Hamiltonian} (see App.~\ref{app:logs} for details). Thus, combining the two contributions, we find the ratio of the splitting to the exchange field $h$ to be:
\begin{equation}
\delta E_0/h = J\rho_0  + (J\rho_0)^2   \ln \left( \frac{D}{\Delta} \right)   + O(J^3).\label{eq:totsplit}
\end{equation}
Fig.~\ref{fig1:nrg-results} shows that this behavior reproduces the splitting in the regime of small $J$ observed in the NRG results. However, the prefactor in the second-order term predicted by the above perturbative treatment is different because the NRG implements a different (lattice) regularization of the impurity Hamiltonian. In other words, the presence of logarithm is a clear indication that the perturbative expression for $\delta E^{(2)}_0/h$ is not universal in the weak coupling limit. In the case of the two-site model, the lack of universality of the perturbative result is manifested in that $\delta E^{(2)}_0/h \propto t$, where $t\sim D$ is a hopping amplitude between the two superconducting sites. Indeed, closer examination reveals that the logarithmic correction in Eqs.~\eqref{eq:2split} and \eqref{eq:totsplit} arises from the non-commutative spin-flip scattering processes that appear in the perturbation theory of the Kondo Hamiltonian and, specifically, from the normal contribution whose Feynman diagram is depicted in Fig.~\ref{fig:feynman}(a). Indeed, using renormalized perturbation theory we could have anticipated the appearance of the logarithm. Let us begin by assuming $\rho_0 J\to 0$, so that the energy splitting is well approximated by the first order result $\delta E_0 = \rho_0 J h$. Under renormalization the exchange coupling constant $g(D) = \rho_0 J(D)$ flows according to~\cite{Hewson_1993}:
\begin{equation}
\bar{D} \frac{d g(\bar{D})}{d \bar{D}} = -g^2(\bar{D}).
\end{equation}
This equation can be integrated from the scale of the initial
bandwidth $D\sim \omega_D$ down to the scale of the gap $\Delta$, where
the renormalization stops. This yields:
\begin{align}
 \rho_0 J(\Delta) &= \frac{1}{\left[\rho_0 J\right]^{-1} 
  - \ln(D/\Delta)}\notag\\
  &=  \rho_0 J + 
 (\rho_0 J)^2 \ln \frac{D}{\Delta} + \cdots \label{eq:regj}
\end{align}
where $J = J(D)$ is the \emph{bare} Kondo coupling at the scale of $D\sim \omega_D$. Provided the coupling remains small, i.e. $\rho_0 J(\Delta) \ll 1$, it is still possible to approximate $\delta E_0/h$ by the first order perturbative result, i.e. $\delta E_0/h = \rho_0 J(\Delta)$. Rewriting $\rho_0 J(\Delta)$ in terms of the bare coupling $\rho_0 J$, we find an infinite series of logarithmic corrections. Furthermore, using this approach we can also remove the unpleasant dependence of $\delta E_0$ on the non-measurable quantities $D$ and $J$ by replacing them with a much more physical energy scale, namely the scale at which the renormalized coupling $g(\bar{D}) = \rho_0 J(\bar{D})$ diverges and defines the Kondo temperature, $T_K =  D\exp\left[-1/(\rho_0 J)\right]$~\cite{Hewson_1993}. Inserting $T_K$ into Eq.~\eqref{eq:regj}, we obtain
\begin{equation}
\rho_0 J(\Delta) = \frac{1}{\ln \left(\frac{\Delta}{T_K} \right)}
\end{equation}
and therefore, provided $T_K \ll \Delta$ (i.e. the weak coupling regime),
\begin{equation}
\delta E_0/h = \rho_0 J(\Delta) =  \frac{1}{\ln \left(\frac{\Delta}{T_K} \right)}.
\end{equation}
In the opposite limit, the strong coupling regime where $T_K \gg \Delta$,   NRG yields $\delta E_0/h \to 2$, independent of $T_K/\Delta$. It is possible to interpolate between the two regimes by writing the ratio $\delta E_0/h$ as a single-parameter scaling function of $T_K/\Delta$, i.e.
\begin{equation}
\delta E_0 = F\left(\frac{T_K}{\Delta} \right),
\end{equation}
where $F(x) \simeq \left(-\ln x \right)^{-1}$ for $x\ll 1$ and $F(x) \to 2$ for $x\to +\infty$.
The shape of this scaling function, as obtained when plotting the  NRG data for $\delta E_0/h$ as a function of $T_K/\Delta$, is shown in Fig.~\ref{fig:scaling}.

Finally, let us emphasize that none of these features are captured by the classical treatment~\cite{yu,shiba,rusinov,skurativska:2023}. Indeed, as we shall describe below, there are also other features of the NRG results that are not captured by the classical treatment.

\section{Other spectral features}\label{sec:spectral}

 Next, we turn our attention to describing the effect of the exchange field of the FMI on the quasi-particle continuum 
 which may be observed in the tunneling spectra. Indeed, NRG shows that other interesting features concern the continuum  threshold  of single-particle excitations, see Fig.~\ref{fig3:NRG-vs-2site}(d, e). To  understand the origin of these features, it is worth recalling the expression of spectral function at positive energies and zero temperature:
 \begin{equation}
 A_{\sigma}(\epsilon > 0) = \sum_{\alpha} |\langle \Phi_{\alpha} | O^{\dag}_{\sigma} | \Phi_0\rangle |^2 \delta\left(\epsilon + E_{\alpha} - E_0\right),\label{eq:speckondo}
 \end{equation}
 where $|\Phi_0\rangle$ is the ground state and $|\Phi_{\alpha}\rangle$ the excited state of the system.
We have introduced the operators $O^{\dag}_{\sigma} = \left[ H_K, c^{\dag}_{0\sigma} \right]$, where $c^{\dag}_{0\sigma} = \sum_{\vec{k}} c^{\dag}_{\vec{k}\sigma}/\sqrt{\Omega}$ creates a fermion with spin $\sigma$ at the position of the impurity. Evaluating the  expressions for $O^{\dag}_{\sigma}$ explicitly, we
 find:
 \begin{align}
 O^{\dag}_{\uparrow} = \frac{J}{2}\left(  S^{+} c^{\dag}_{0,\downarrow} + S^{z} c^{\dag}_{0\uparrow} \right),\\
  O^{\dag}_{\downarrow} = \frac{J}{2}\left(  S^{-} c^{\dag}_{0,\uparrow} - S^{z} c^{\dag}_{0\downarrow} \right).
 \end{align}
 When the spin is treated as a classical vector, Refs.~\cite{yu,shiba,rusinov}, the terms proportional to $S^{\pm}$ are neglected and  the above operators are approximated as $O_{\sigma} \approx \pm S^z c_{0\sigma}$ (where $+$ is for $\sigma = \uparrow$ and $-$ for $\sigma = \downarrow$)~\footnote{This approximation can be justified in the large spin $S$ limit with $S^z = O(S)$ and $S^{\pm} = O(\sqrt{S})$.}. Thus, $A_{\sigma}(\epsilon)$ reduces to the imaginary part of the Fourier transform of the local (retarded) Green's function, $G^R_{\sigma}(t) = -i \theta(t) \langle \{c_{0\sigma}(t), c^{\dag}_{0\sigma}(0) \} \rangle$. The calculation
 of the spectral function for a magnetic impurity coupled to a spin-split superconductor in the classical approximation was reported in Ref.~\cite{skurativska:2023}. It  yields spin-dependent  threshold energies, $\epsilon_{B\sigma}  = \Delta \pm h$, 
 for the continuum of single-particle excitations that does not differ from the one obtained from the
 unperturbed Green's functions of the superconductor (cf. \ref{eq:gpm} in Appendix),   independently of whether the impurity is in the strong or weak coupling regime.

 On the other hand, NRG yields a very different picture, see Fig~\ref{fig3:NRG-vs-2site}(d, e). In the strong coupling limit, the value of the threshold energy is $\epsilon_{B\sigma} = \Delta\pm h$ and agrees with the result obtained from the classical approximation. However, in the weak coupling where the classical approximation is expected to be more accurate, NRG shows  that $\epsilon_{B\uparrow}$ undergoes a  downward shift from its ``classical'' value $\Delta + h$ and becomes almost equal to $\epsilon_{B\downarrow} \simeq \Delta - h$ (see Fig.~\ref{fig3:NRG-vs-2site}d). To better understand the origin of this shift in $\epsilon_{B\uparrow}$, let us first recall that for $h > 0$, the ground state in the weak coupling regime is an eigenstate of $S^z_T$
 with eigenvalue $m = -\tfrac{1}{2}$, i.e. $S^z_T | \Phi_0\rangle = \left(-\tfrac{1}{2}\right)|\Phi_0\rangle$. As pointed out above, this state is to a large extent well approximated by the product state $|\text{BCS}\rangle |-\tfrac{1}{2}\rangle$. Mathematically,
 this is expressed by the following equation:
 \begin{equation}
 |\Phi_0\rangle = Z^{1/2} |\text{BCS}\rangle |-\tfrac{1}{2}\rangle + \cdots 
 \end{equation}
where $Z\lesssim 1$ and the ellipsis stays for other components of the state containing e.g. quasi-particle triplet excitations similar to those discussed in Sec.~\ref{sec:two-sitepert} for the two-site model.
Using this insight, we notice that for $h > 0$, the action of the operators $O^{\dag}_{\sigma}$ on the ground state produces quite different results:
\begin{align}
O^{\dag}_{\uparrow} |\Phi_0\rangle &=   \frac{J}{2}   \big( Z^{1/2}  c^{\dag}_{0\downarrow} | \text{BCS}\rangle |+\tfrac{1}{2}\rangle  \notag\\ 
&\qquad -  \tfrac{1}{2} Z^{1/2}  c^{\dag}_{0\uparrow} |\text{BCS}\rangle  |-\tfrac{1}{2}\rangle \big)   
+  \cdots \label{eq:spinflip}\\
O^{\dag}_{\downarrow} |\Phi_0\rangle &=   \frac{J}{2} Z^{1/2} c^{\dag}_{0\downarrow} |\text{BCS}\rangle |-\tfrac{1}{2}\rangle +
\cdots \label{eq:nospinflip}
\end{align}
Eq.~\eqref{eq:spinflip} shows that when an electron with spin anti-aligned with the ground state projection of $S^z_T$ tunnels it can flip in the impurity spin. However, no such spin flip is produced when
the spin is aligned with the $S^z_T$ projection of the weak-coupling ground state. The spin-flip has important consequences for the single-particle excitations created in the superconductor by  $c_{0\downarrow}$: By flipping the spin of the impurity from $|-\tfrac{1}{2}\rangle$ to $|+\tfrac{1}{2}\rangle$, the Ising term of $H_K$, that is, $J S^z s^z_0$,  changes from repulsive to attractive for single-particle excitations with $\sigma = \downarrow$. In other words, some of the excitations created by $S^{+} c^{\dag}_{0\downarrow}$ live in a sector of the Hilbert space where the Ising term has the opposite sign to those excitations created by $S^z c^{\dag}_{0\uparrow}$.
The attractive character of the Ising term induces an ``excitonic-like''~\cite{mah00} shift of the excitation energies and therefore the threshold energy for the continuum $\epsilon_{B\uparrow}$ becomes lower than the classical value $\Delta + h$. 
Let us point out that for $h = 0$ no such shift occurs because the ground state is a degenerate doublet and the spectral function must be averaged over
the two states of the doublet~\cite{skurativska:2023}. Thus, the spin-flip contribution is equal to zero by time-reversal symmetry. Similarly, for $h\neq 0$ in the strong coupling regime, the ground state is a singlet and therefore contains components of both spin orientations of the impurity, which are strongly entangled with a cloud of quasi-particle excitations in the superconductor. Since   the spin projection of the impurity rapidly fluctuates in a singlet, the Ising term of $H_K$ should not affect
the energy of single-particle excitations. 
 
 Finally, let us point out that the effects described above are also qualitatively captured by the 
two-site model introduced in Sec.~\ref{sec:2site}.  For this model, we can also compute the spectral function using the same operators $O_{\sigma}$ as
in the NRG case. For $J< J_{\text{cr}}$ we find that the spin-down component of the continuum part of the spectral function shifts by $-h-\delta h$, while the spin-up component shifts by a different amount of $-h$. Fig. \ref{fig3:NRG-vs-2site}) (b,c) shows this different behavior of the spin-up and spin-down components of the spectral function in the weak and strong coupling regimes of $J$.

\section{conclusions}\label{sec:fin}

We have investigated a system of a single magnetic impurity on the surface of a ferromagnetic insulator (FMI)/superconductor (SC) heterostructure, where the SC has a thickness smaller than the coherence length. Under such conditions, the proximity effect of the FMI on the SC and impurity is twofold: First, it splits the energy of the Bolgoliubov quasi-particle states without destroying the superconducting state. Second, it leads to a spin polarization of the surface of the superconductor that contains the impurity which results in an RKKY-like interaction between the latter and the FMI. We have studied this system by applying the numerical renormalization group method (NRG).

From the NRG calculations, we obtained the ground-state and low-energy excitations of the system as a function of the (Kondo) exchange of the impurity. The latter drives a quantum phase transition from a unique even-parity ground state to a (unique) singlet ground state.
Our main result is the energy splitting of the parity-even doublet with total spin $\pm \tfrac{1}{2}$ induced by the magnetic proximity effect.  The splitting occurs already in the regime of weak exchange coupling $J$ and grows smoothly with $J$ saturating in the strong coupling limit to twice the value of the exchange field with the FMI. In the strong coupling regime, the splitting is observable in the spectra acquired using the scanning tunneling microscope as the splitting of the YSR excitations. 

In the limit of weak Kondo coupling $J$, we have  computed the splitting using
perturbation theory. The splitting contains a first-order correction in $J$ from the RKKY-like interaction and a second-order correction from the non-commutative Kondo-like scattering with the host electrons. The second-order correction is of the form of a typical Kondo logarithm, which is cut off by the superconducting gap $\Delta$. This result has been also re-derived using scaling,
and when combined with the numerical observation of the saturation at large $J$, it allows us to conjecture that the ratio of the splitting to the FMI exchange field can be written as a single-parameter scaling function of $T_K/\Delta$, where $T_K$ is
the Kondo temperature of the impurity. Additionally, we have also discussed the shift of the threshold for the continuum of single-particle excitations observed in the spin-resolved spectral function computed using NRG. The shift occurs in the weak coupling regime only
for tunneling electrons with the spin opposite to the spin of the impurity in the ground state
and has been qualitatively explained as an excitonic-like effect. 
Together with the splitting of the even-parity ground state doublet, this effect cannot
be captured by the classical approach of Yu, Shiba, and Rusinov~\cite{yu,shiba,rusinov} when generalized to the quantum impurity model in Ref.~\cite{skurativska:2023}.

For a qualitative understanding of the underlying physics of this system, we have employed a two-site model, where the superconductor is represented by two spinful fermion sites with the impurity spin coupled to one of them. This minimal model correctly captures the behavior of energy splitting of the even-parity doublet a  and provides an intuitive picture in terms of coupling to triplet excitations to explain the physics behind it. It also qualitatively explains the shift of the threshold referred to above. 

The splitting is proportional to the exchange field of the FMI and it can be quite sizable for a sufficiently large Kondo coupling. It therefore allows using the FMI to control the spin of the impurity and its excitations \emph{in the absence} of an external magnetic field. This happens despite that the FMI is located rather far (typically tens of nm) from the impurity. It was already pointed out in Ref.~\cite{skurativska:2023} that the exchange field of the FMI leads to robust spin polarization of the YSR states. This effect is further enhanced by the magnetic and triplet correlations discussed in this work. 
Besides the STM in the strong coupling regime, where the splitting of the doublet is observable as the splitting of the YSR peaks for the two spin orientations, the splitting may also be observable by measuring the microwave absorption of a dilute ensemble of
impurities on the surface of the FMI/SC.

\acknowledgments 
MAC and SB thank Ilya Tokatly for useful discussions on the magnetic proximity effect.
This work has been supported by the Agencia Estatal de Investigación del Ministerio de Ciencia e Innovación (Spain) through Grants No. PID2020-120614GB-I00/AEI/10.13039/501100011033 (ENACT),  PID2020-114252GB-I00/AEI/10.13039/501100011033, and TED2021-130292B-C42.
C.-H.H. acknowledges support from a Ph.D. fellowship from DIPC. A. S. acknowledges funding from the Basque Government’s IKUR initiative on Quantum technologies (Department of Education).

\appendix

\section{Spin-splitting in perturbation theory}\label{app:split}
The Hamiltonian describing a magnetic impurity in the spin-split superconductor
has been introduced in Sec.~\ref{sec:model}. In this Appendix, to make the
notation more compact, we shall rewrite it as follows:
\begin{align}
&H = H_0 + H_I,\\
&H_I = J \vec{S}\cdot \vec{s}_0+ J\rho_0 h S^z,\label{eq:hk} 
\end{align}
where $H_0$ describes the clean spin-split superconductor (cf. Eq.~\ref{eq:h0}) and $H_I = H_K + H_{RKKY}$ describes all the interactions of the impurity with its host; $\vec{s}_0$ is the spin-density operator of the superconductor at the location of the impurity;  $\vec{S}$ is the impurity spin operator. 

Let us compute the free energy in perturbation theory. We start with the following perturbation theory
result for the shift of the grand-canonical free energy at absolute temperature $T$:
\begin{align}
\Delta F^{m} &= F^m - F_0 \notag \\
&= -T \ln \left\{  \langle m| \langle  \mathcal{T} \exp\left[ - \int^{1/T}_0 d\tau\, H_I(\tau)\right]  \rangle | m \rangle \right\}\notag \\
&= T\int^{1/T}_0 d\tau_1 \,  \langle m | \langle     H_I(\tau_1)   \rangle |m\rangle\notag \\
&-\frac{T}{2} \int^{1/T}_0 d\tau_1 d\tau_2\,  \langle m | \langle \mathcal{T} \left[ H_I(\tau_1) H_K(\tau_2) \right]\rangle |m\rangle \notag\\
&\qquad\qquad + O(J^3).
\end{align}
In the above expression $\langle \ldots \rangle$ stands for the average over the canonical ensemble of the eigenstates of the superconductor Hamiltonian, $H_0$ and $|m  =\pm\tfrac{1}{2}\rangle$ are the eigenstates of decoupled magnetic impurity with $S^z  = \pm\tfrac{1}{2}$. In this basis, the first order term vanishes because $\langle \vec{s}_0\rangle = 0$. Note that we do not allow the state of the impurity to fluctuate thermally because we are interested in computing the ground-state energy difference between the $m=+\tfrac{1}{2}$ and $m=-\tfrac{1}{2}$ configurations. Since the unperturbed ground state is the product state of the BCS wave function and a single impurity spin, the first-order contribution of $H_K$ vanishes. The first-order term is simply given by the RKKY contribution:
\begin{align}
    \Delta F^{(1,m) } = J\rho_0 h m \,,
\end{align}
that leads to the splitting in energy,
\begin{align}
    \Delta E^{(1) }_0 = J\rho_0 h\,.
\end{align}
Introducing Eq.~\eqref{eq:hk}, we obtain  the second order contribution:
\begin{align}
\Delta F^{(2,m)} &=  -\frac{J^2 T}{2} \int^{1/T}_0 d\tau_1d\tau_2 \langle m| \mathcal{T} \left[ S^a(\tau_1) S^b(\tau_2) \right] | m\rangle \notag\\  &\times\langle \mathcal{T} \left\{ \left[\rho_0 h \delta_{a,z}+ s^{a}_0(\tau_1) \right]\left[\rho_0 h \delta_{b,z}+s^{b}_0(\tau_2)\right] \right\} \rangle  \notag\\
&+O(J^3).
\end{align}
Next, we compute the impurity spin correlation function. Assuming $\vec{S}$ to be a spin-$\tfrac{1}{2}$  operator, we have
\begin{multline}
\langle m| \mathcal{T} \left[ S^a(\tau_1) S^b(\tau_2) \right] | m\rangle = \theta(\tau_1-\tau_2) \langle m| S^a  S^b | m\rangle  \notag\\
+ \theta(\tau_2-\tau_1)  \langle m|  
S^b  S^a | m\rangle \notag\\
=  \tfrac{1}{4}  \delta^{ab} + \tfrac{i}{2} \epsilon^{abc}\langle m| S^c | m\rangle 
\times \, \mathrm{sgn}(\tau_1-\tau_2),
\end{multline}
where used the Pauli-matrix identity  $S^a S^b = \tfrac{1}{4} \delta^{ab} \mathbb{I} + \tfrac{i}{2} \epsilon^{abc} S^c$.  Note that $\langle m | S^c | m\rangle = \delta^{z,c} m$, because $S^z$ is the only diagonal spin operator in the basis $|m\rangle$. Hence, since, we are interested only in the difference between $\Delta F^{(2,m=+\tfrac{1}{2}}$ and $\Delta F^{(2,m=-\tfrac{1}{2})}$, the splitting of the free energy is given by
\begin{multline}
\Delta F^{(2)}_{0} = \Delta F^{(2,m=+\tfrac{1}{2})} - \Delta F^{(2,m=-\tfrac{1}{2})} \\
\simeq \frac{i  J^2 T}{4}  \int^{1/T}_0  d\tau_1 d\tau_2  \,  \mathcal{C}^z(\tau_1-\tau_2) \mathrm{sgn}(\tau_1-\tau_2).
\end{multline}
In the last line, we have introduced the correlation function 
\begin{align}
\mathcal{C}^z(\tau) &= \epsilon^{ab,z}  \mathcal{C}^{ab}(\tau_1-\tau_2),\\
\mathcal{C}^{ab}(\tau_1-\tau_2) &=- \langle \mathcal{T} \left[ s^a_0(\tau_1) s^b_0(\tau_2)\right] \rangle
\end{align}
Next, let us introduce the Fourier transform of the above spin correlation function:
\begin{align}
\mathcal{C}^{ab}(\tau_1-\tau_2)  = T \sum_{\omega_n} e^{-i\omega_n(\tau_1-\tau_2)} \mathcal{C}^{ab}(i\omega_n),
\end{align}
which yields
\begin{equation}
\Delta F^{(2)}_0 = \frac{J^2 T}{2}  \sum_{\omega_n} \frac{\mathcal{C}^{z}(i\omega_n)}{\omega_n}
\end{equation}
In the last line we have used that ($\omega_n = 2\pi n T$):
\begin{equation}
\int^{1/T}_0 d\tau_1 d\tau_2
\, \mathrm{sgn}(\tau_1-\tau_2) e^{-i\omega_n(\tau_1-\tau_2)} = -\frac{2i}{T \omega_n}
\end{equation}
In addition, using $S^{\pm} = S^x \pm i S^y$ we have
\begin{equation}
\mathcal{C}^{z}(\omega_n) = \frac{i}{2} \left[ \mathcal{C}^{+-}(i\omega_n) -\mathcal{C}^{-+}(\omega_n) \right],
\end{equation}
where we have used that $\mathcal{C}^{++}(\omega_n) = \mathcal{C}^{--}(\omega_n) = 0$ because of conservation of total spin $z$-projection, i.e. $S^z_T$. Therefore,
\begin{equation}
\Delta F^{(2)}_{0} = \frac{ i J^2T}{4} \sum_{\omega_n} \frac{\mathcal{C}^{+-}_0(i\omega_n)-\mathcal{C}^{-+}_0(i\omega_n)}{\omega_n} + O(J^3).
 \end{equation}
Taking the $T\to 0$ limit, the above sum becomes an integral, which equals the ground energy splitting:
\begin{equation}
\delta E^{(2)}_0 = -\frac{J^2}{4} \int \frac{d\omega}{2\pi} \,\left[  \frac{\mathcal{C}^{+-}(i\omega)  - \mathcal{C}^{-+}(i\omega)}{i\omega}  \right] + O(J^3). \label{eq:intres0}
\end{equation}
Finally, recalling that at $T = 0$,
\begin{equation}
\int^{+\infty}_0 d\tau \, \mathcal{C}^{ab}(\tau) =   \int \frac{d\omega}{2\pi} \frac{\mathcal{C}^{ab} (\omega)}{i\omega}
 \end{equation}
we arrive at
\begin{equation}
\delta E^{(2)}_0 = -\frac{J^2}{4} \int^{+\infty}_{0} d\tau \left[ \mathcal{C}^{+-}(\tau) - \mathcal{C}^{-+}(\tau)\right] + O(J^3). \label{eq:intres} 
\end{equation}
Indeed, using the spectral representation of the spin correlation functions $\mathcal{C}^{ab}(\tau)$ at $T = 0$, i.e.
\begin{align}
\chi^{ab} &= - \int^{+\infty}_0 d\tau\, \mathcal{C}^{ab}(\tau) = \int^{+\infty}_0 d\tau\, \langle\mathcal{T}\left[ s^a_0(\tau) s^b_0(0)\right] \rangle \notag\\
&= \sum_{E} \frac{\langle \text{BCS} | s^a_0 | E \rangle
\langle E| s^b_0 | \text{BCS} \rangle}{E-E_0},
\end{align}
we can rewrite Eq.~\eqref{eq:intres} as the difference:
\begin{equation}
\Delta E^{(2)}_{0} = \frac{J^2}{4}\left[ \chi^{+-}-\chi^{-+}\right]
\end{equation}
Indeed, a quicker way to arrive at this expression is to start from the second-order perturbation theory formula at $T = 0$:
\begin{widetext}
\begin{align}
\Delta E^{(2)}_{m=1/2} &= \sum_{E, n=\pm \tfrac{1}{2}}  \frac{|\langle n| \langle E | H_K | \text{BCS}\rangle |m=+1/2|^2}{E_0-E}\notag\\
&= -J^2 \sum_{E}  \left[   \frac{1}{4} \frac{\langle \text{BCS} | s^{-} | E\rangle \langle E | s^{+} | \text{BCS}\rangle}{E-E_0} \sum_{n=\pm \tfrac{1}{2}} \langle m =+\tfrac{1}{2} | S^{+}| n\rangle \langle n | S^{-} | m=+\tfrac{1}{2}\rangle  \right]\notag  \\ 
&- J^2 \sum_E \left[ \frac{\langle \text{BCS} | s^{z} | E\rangle \langle E | s^{z} | \text{BCS}\rangle}{E-E_0}  \sum_{n=\pm \tfrac{1}{2}} \langle m =+\tfrac{1}{2} | S^z n\rangle \langle n | S^{z} | m=+\tfrac{1}{2}\rangle   \right] \\
&= -J^2 \sum_{E}  \left[   \frac{1}{4} \frac{\langle \text{BCS} | s^{-} | E\rangle \langle E | s^{+} | \text{BCS}\rangle}{E-E_0} \langle m =+\tfrac{1}{2} | S^+ S^{-} | m=+\tfrac{1}{2}\rangle  \right]  \notag\\ 
&- J^2 \sum_E \left[  \frac{\langle \text{BCS} | s^z | E\rangle \langle E | s^{z} | \text{BCS}\rangle}{E-E_0}  \langle m=+\tfrac{1}{2} | (S^z)^2 |m=+\tfrac{1}{2}\rangle \right].
\end{align}
The last term involving $(S^z)^2$ is the same for the other ground state of the doublet with $m = -\tfrac{1}{2}$ and
does not contribute to the splitting. Therefore, 
\begin{align}
\Delta E^{(2)}_0 &= -\frac{J^2}{4} \sum_{E}  \left[    \frac{\langle \text{BCS} | s^{-} | E\rangle \langle E | s^{+} | \text{BCS}\rangle}{E-E_0} \langle m =+\tfrac{1}{2} | S^+ S^{-} | m=+\tfrac{1}{2}\rangle  \right]   \notag\\
 & \qquad + \frac{J^2}{4} \sum_{E} \left[  \frac{\langle \text{BCS} | s^{+} | E\rangle 
 \langle E | s^{-} | \text{BCS}\rangle}{E-E_0} \langle m =-\tfrac{1}{2} | S^{-} S^{+} | m=-\tfrac{1}{2}\rangle  \right] \\
 &= \frac{J^2}{4}\left(\chi^{+-}-\chi^{-+} \right).
\end{align}
where we have used that $\langle m =-\tfrac{1}{2} | S^{-} S^{+} | m=-\tfrac{1}{2}\rangle   = 1$,
etc.
\end{widetext}

\section{Dependence on triplet correlations}\label{app:tripletcorr}
In this section, we evaluate the energy splitting as given by Eq.~\eqref{eq:intres0} by computing the correlation functions  $C^{+-}(i\omega)$ and $C^{-+}(i\omega)$. To this end, we recall that in terms of
the (four-component) Nambu spinor 
\begin{equation}
\Psi_0 =  \left( 
\begin{array}{c}
c_{0\uparrow} \\
c_{0\downarrow}\\
-c^{\dag}_{0\downarrow}\\
c^{\dag}_{0\uparrow}
\end{array}
\right)= 
\left(
\begin{array}{c}
C_{0}\\
C^{\dag}_0 i\sigma^2 
\end{array}
\right) 
\end{equation}
the local spin operator reads $\vec{s}_0  =\tfrac{1}{2} \Psi^{\dag}_0 \vec{s}\tau^0 \Psi_{0}$.
Thus, using this notation we can write the correlation functions of interest as follows:
\begin{align}
\mathcal{C}^{ab}(\tau) &= -\langle\mathcal{T} \left[ s^a(\tau) s^b(0) \right] \rangle \notag\\
 &= \frac{1}{2}\mathrm{Tr} \left[ \mathcal{G}^{0}(-\tau) s^a\tau_0 \mathcal{G}^{0}(\tau) s^b\tau^0\right]
\end{align}
To obtain the expression in the last line, we have employed Wick's theorem and
re-writen $\mathcal{C}^{ab}(\tau)$ in terms of the fermion local Green's function $\mathcal{G}^0(\tau)$
(see further below for the explicit expression of the latter). Next, upon performing the Fourier transform at $T= 0$ we arrive at:
\begin{align}
\mathcal{C}^{ab}(\omega) &= \int d\tau \, e^{i\omega\tau} \mathcal{C}^{ab}(\tau) \notag\\
&=\frac{1}{2} \int \frac{d\omega^{\prime}}{2\pi} 
\mathrm{Tr} \left[ \mathcal{G}^{0}(i\omega^{\prime}) s^a\tau_0 \mathcal{G}^{0}(i\omega-i\omega^{\prime}) s^b\tau^0\right]. 
\end{align}
In the last line, we have relabelled $\omega_1$ as $\omega^{\prime}$. 

It is interesting to write the local Green's function as follows:
\begin{align}
\mathcal{G}^0(i\omega) &= g^0_s(i  \omega) s^0 + g^{0}_t( i\omega)   s^z, \label{eq:g0st}\\
g^0_s(i\omega) &= \frac{1}{2} \left[ \mathcal{G}^0_{+}(i\omega) + \mathcal{G}^0_{-}(i\omega)\right],\notag\\
g^0_t(i\omega) &= \left[ \mathcal{G}^0_{+}(i\omega) - \mathcal{G}^0_{-}(i\omega)\right].\notag
\end{align}
where (in the wide band limit)
\begin{align}
\mathcal{G}^0_{\pm}(i\omega) = -\pi \rho_0 \frac{(i\omega \mp h) \tau^0 + \Delta\tau^x}{\sqrt{\Delta^2-(i\omega \mp h)^2}}. \label{eq:gpm}
\end{align}
Hence,
\begin{align}
\mathcal{C}^{ab}(i\omega) 
&= \frac{1}{2} \left\{ \mathrm{Tr} \left[ \left( g_s\star g_s - g_t\star g_t \right)(i\omega) \otimes s^a s^b \right] \right. \notag\\ 
&\left. \qquad +  
\mathrm{Tr} \left[ \left( g_t\star g_s - g_s \star g_t\right)(i\omega) \otimes s^z  s^a  s^b \right]   \right\}.
\end{align}
In the above expression, we have used the notation:
\begin{equation}
(f \star g)(i\omega) = \int \frac{d\omega^{\prime}}{2\pi} f(i \omega^{\prime}) g(i\omega - i\omega^{\prime}) 
\end{equation}
for the convolution of two functions $f(i\omega)$ and $g(i\omega)$ of the Matsubara frequency $i\omega$;
$\otimes$ stands for the Kronecker product of the matrices in the (particle-hole) Nambu and spin indices.  The energy shift is determined by the difference:
\begin{multline}
\mathcal{C}^{+-}(i\omega) - C^{-+}(i\omega) =  \frac{1}{2}\mathrm{Tr} \left( g_t \star g_s - g_s\star g_t\right)(i\omega). 
\end{multline}
In the last line we have used that $\mathrm{Tr} \left[ m\otimes s^z \right] = 0$ and $\mathrm{Tr} \left[ m\otimes s^0 \right]  = 2 \: \mathrm{Tr} \: (m)$ for any $2\times 2$ matrices in Nambu indices.
Notice that  $g_s$ and $g_t$ are $2\times 2$ matrices in the (particle-hole) Nambu indices. We have also used that $s^a,s^b \in \{s^{+},s^{-}\}$ and therefore anti-commute with $s^z$. 
Hence, 
\begin{align}
\chi^{+-}-\chi^{-+} &=  - \int \frac{d\omega}{2\pi} \frac{\mathcal{C}^{+-}(i \omega) - \mathcal{C}^{-+}(i\omega)}{i\omega} \notag\\
&= \frac{1}{2}\int \frac{d\omega}{2\pi} \frac{\mathrm{Tr} \left[ g_s\star g_t -  g_t \star g_s\right](i\omega)}{i\omega}\notag\\
&=  \int \frac{d\omega}{2\pi}\,  
\frac{\mathrm{Tr}  \left[ g_s\star g_t\right] (i\omega) }{i\omega}.\label{eq:result}
\end{align}
In the last line, we have used the identity:
\begin{align}
\mathrm{Tr} \left[ f \star g \right](i\omega) &= \int \frac{d\omega^{\prime}}{2\pi} \mathrm{Tr} \: f(i \omega^{\prime}) g(i\omega - i\omega^{\prime}) \\
&= \int \frac{d\omega^{\prime\prime}}{2\pi} \mathrm{Tr} f(i\omega^{\prime\prime}+i\omega) g(i\omega^{\prime\prime})\notag\\
 &=  \int \frac{d\omega^{\prime\prime}}{2\pi} \mathrm{Tr} \: g(i\omega^{\prime\prime}) f(i \omega^{\prime\prime}+i\omega)\notag \\
 &= \mathrm{Tr}\left[ g\star f\right](-i\omega).\label{eq:identity}
\end{align}
%

%
%
Thus, we arrive at the following expression for the energy splitting to second order in $J$:
\begin{align}
\delta E_0 &= J h\rho_0 h+ \frac{J^2}{4}\left(\chi^{+-}-\chi^{-+}\right)\notag \\
&=J h\rho_0 + \frac{J^2}{4} \int \frac{d\omega}{2\pi}  \frac{\mathrm{Tr}\left[g_s\star g_t\right](i\omega)}{i\omega}.
\end{align}
As discussed in the main text,  this expression vanishes as $h\to 0$ due to the
vanishing (odd-frequency) triplet correlations described by $g_t\propto h\to 0$. 

\section{Logarithmic divergence}\label{app:logs}

In this Appendix, we explicitly show that the expression for the energy splitting depends logarithmically on the ratio of the bandwidth to the gap. It turns out this can be most
conveniently shown by performing the integral over imaginary 
time $\tau$ using Eq.~\eqref{eq:intres}. Let us recall that
$s^{+}_0 =  c^{\dag}_{0\uparrow} c_{0\downarrow}$ and $s^{-} = c^{\dag}_{0\downarrow} c_{0\uparrow}$. 
%
%
%
For the calculation of the correlation functions 
\begin{align}
\mathcal{C}^{+-}(\tau) &= -\langle \mathcal{T} \left[s^{+}_0(\tau) s^{-}_0(0) \right] \rangle,\notag \\  
\mathcal{C}^{-+}(\tau) &=- \langle \mathcal{T} \left[s^{-}_0(\tau) s^{+}_0(0) \right] \rangle \\
\label{eq:cmp}
\end{align}
using Wick's theorem we need the local correlation functions for spin up and down single-particle excitations, whose Fourier transform is displayed in Eq.~\eqref{eq:gpm}. Computing their inverse Fourier transforms in imaginary time for $T = 0$, we find 
\begin{align}
\mathcal{G}^0_{\alpha_{\sigma}}(\tau) &= - \langle \mathcal{T} \left[
\left(
\begin{array}{c}
c_{0 \sigma}(\tau) \\ 
c^{\dag}_{0,-\sigma}(\tau)
\end{array}
\right)\otimes
\left(
\begin{array}{cc}
c^{\dag}_{0\sigma}(0) &
c_{0,-\sigma}(0)
\end{array}
\right)
\right] \rangle \notag\\
&=-\pi \rho_0 \int \frac{d\omega}{2\pi} e^{-i\omega\tau } \frac{\left[(i\omega -\alpha h) \tau^0 + \Delta \tau^x \right]}{\sqrt{\Delta^2-(i\omega - \alpha_{\sigma} h)^2}} \notag \\
&= -\rho_0 \Delta  e^{-\alpha_{\sigma} h\tau} \left[  \mathrm{sgn}(\tau) K_1(\Delta |\tau|) \tau^0 + \right.  \notag\\
 &\left.\qquad + K_0(\Delta |\tau|)  \tau^x \right],
\label{eq:lg}
\end{align}
where $\alpha_{\uparrow} = +1$ and $\alpha_{\downarrow}  = -1$.
The last expression in terms of the Bessel $K_0$ and $K_1$ functions is valid in the wideband limit where 
$|\tau|\gg \tau_c$, where $\tau_c \sim D^{-1}$,
 $D$ being the bandwidth. Hence, we obtain
%
%
and introducing the result of Eq.~\eqref{eq:lg} one obtains:
\begin{align}
C^{+-}(\tau) &= - \rho^2_0 \Delta^2 e^{+2h\tau}\left[ K^2_1(\Delta |\tau|) -  K^2_0(\Delta |\tau|) \right],\\
C^{-+}(\tau) &= - \rho^2_0  \Delta^2   e^{-2h\tau}\left[ K^2_1(\Delta |\tau|) -  K^2_0(\Delta |\tau|) \right].
\end{align}
Therefore, for $|\tau| \gg \tau_c$, 
\begin{align}
C^{+-}(\tau) - C^{-+}(\tau) &= -2 \rho^2_0  \Delta^2  \left[ K^2_1(\Delta |\tau|) -  K^2_0(\Delta |\tau|) \right] \notag \\
&\times  \sinh(2h \tau)
\end{align}
and, introducing the cut off $\tau_c$ at short times, we have
\begin{align}
\chi^{+-}-\chi^{-+} \simeq - \int^{+\infty}_{\tau_c} d\tau \, \left[ C^{+-}(\tau) - C^{-+}(\tau) \right]\notag \\
= 2\rho^2_0 \Delta^2   \int^{+\infty}_{\tau_c} d\tau \, \left[K^2_1(\Delta \tau) -  K^2_0(\Delta \tau)  \right]  \sinh(2 h\tau).
\end{align}
Note that this expression makes sense only if $|h| <\Delta$, which is the requirement for 
stability of the spin-split superconductor. 
Next, we show that this expression is logarithmic divergent in the limit where $\tau_c\to 0$. To see this, 
let us recall that for small argument $u = \Delta \tau$,
\begin{align}
K_0(u) &=  -\ln(u/2)  -\gamma  + O(u^2 \ln(u)),\\
K_1(u) &= \frac{1}{u} + O(u \ln(u)),
\end{align}
where $\gamma$ is Euler's constant. 
Thus, we see that the term involving $K^2_1(\Delta \tau)$ is most singular and behaves as 
$\sim \tau^{-1}$ as $\tau \to 0$, which means that the above integral
\begin{align}
\chi^{+-}-\chi^{-+} &\simeq  \rho^2_0 \Delta^2  \int^{1/\Delta}_{\tau_c} d\tau\: \frac{4h}{\Delta^2 \tau} + \mathrm{regular \, terms}\notag\\
&= 4\rho^2_0  h \ln \left( \frac{D}{c\Delta} \right) + \mathrm{regular \, terms},
\end{align}
where we have set $\tau_c  = c/D$, where $c \sim 1$ is a constant to be specified below.
The (regular) contribution from the anomalous diagram in the limit where $h/\Delta\ll 1$ can be expressed
in terms of the integral:
\begin{align}
&\int^{+\infty}_{\tau_c \Delta} du K^2_0(u) \sinh\left( \frac{2 h}{\Delta} u \right)\notag \\
&=  \frac{2h}{\Delta} \int^{+\infty}_{\tau_c\Delta} d u \: u K^2_0(u) + O\left( \frac{h^2}{\Delta^2}\right) \notag\\
&= \frac{h}{\Delta} + O\left( \frac{h^2}{\Delta^2}\right)
\end{align}
where we have taken the (wide band) limit where $\tau_c \Delta \sim \Delta/D \to 0$
and used $\int^{+\infty}_0 du \: u K^2_0(u) = \frac{1}{2}$.

Thus, for small $h$, we find that the energy shift takes the form:
\begin{equation}
\delta E_0 = J\rho_0 h+ (J\rho_0)^2 h \left[ \ln \left( \frac{D}{c \Delta} \right) - \frac{1}{2} \right] + O(J^3,h^2).
\end{equation}
We can set the freedom to define the imaginary time cut-off $\tau_c$ in terms of $D^{-1}$ and choose $c  = e^{-1/2}$ in order to absorb the anomalous contribution into the logarithm, which yields the following results quoted in the main text:
\begin{equation}
\delta E_0/h = J\rho_0 + (J\rho_0)^2  \ln \left( \frac{D}{\Delta} \right)  + O(J^3,h).
\end{equation}

\section{Perturbation theory for the two-site model}
\label{app:B}

For $J=0$ the ground state of a two-site model can be obtained by diagonalizing the following 
two-site quadratic Hamiltonian:
\begin{align}
    H_0 &= \sum_{j=0,1} \left[ \Delta \left(c^{\dag}_{j\uparrow} c^{\dag}_{j\downarrow} + \text{H.c.}\right) 
    - h \left(c^{\dag}_{j\uparrow} c_{j\uparrow} - c^{\dag}_{j\downarrow} c_{j\downarrow}\right) 
    \right] \notag \\
    &\qquad - t \sum_{\sigma} \left(c^{\dag}_{0\sigma} c_{1\sigma} + \text{H.c.}\right).
\end{align}
To this end, it is convenient to rewrite Hamiltonian in the basis of bonding and anti-bonding orbitals described by the operators: 
\begin{equation}
c_{\pm,\sigma} = \frac{1}{\sqrt{2}}(c_{0\sigma} \pm c_{1\sigma}),
\end{equation}
thus, 
\begin{align}
    H_0 &= \sum_{l= \pm} \Delta (c^\dagger_{l\uparrow} c^\dagger_{l\downarrow} + h.c.) -h (c^\dagger_{l\uparrow} c^{ }_{l\uparrow} - c^\dagger_{l\downarrow} c^{ }_{l\downarrow}) \notag\\ 
    &\qquad - t \sum_{\sigma = \uparrow, \downarrow} (c^\dagger_{+\sigma} c^{ }_{+\sigma} - c^\dagger_{-\sigma} c^{ }_{-\sigma}). 
\end{align}
This Hamiltonian can be diagonalized through the following Bogoliubov transformation:
\begin{equation}
\begin{split}
    \gamma_{\pm\uparrow} &= u_\pm c_{\pm\uparrow} - v_\pm c^\dagger_{\pm\downarrow}\,, \\
    \gamma_{\pm\downarrow} &= u_\pm c_{\pm\downarrow} + v_\pm c^\dagger_{\pm\uparrow}\,,
    \end{split}
\end{equation}
where 

\begin{align}
\label{appeq:u-vpar}
u_+ &= \frac{t + \sqrt{t^2 + \Delta^2}}{\Delta \sqrt{1 + \left(\frac{t + \sqrt{t^2 + \Delta^2}}{\Delta}\right)^2}} \,,\\
v_+ &= -\frac{1}{\sqrt{1 + \left(\frac{t + \sqrt{t^2 + \Delta^2}}{\Delta}\right)^2}} \,,\\
u_- &= \frac{-t + \sqrt{t^2 + \Delta^2}}{\Delta \sqrt{1 + \left(\frac{-t + \sqrt{t^2 + \Delta^2}}{\Delta}\right)^2}} \,, \\
v_- &= -\frac{1}{\sqrt{1 + \left(\frac{-t + \sqrt{t^2 + \Delta^2}}{\Delta}\right)^2}}\,.
\end{align}
The ground state is given by
\begin{equation}
    |\text{GS}_m\rangle = |\text{BCS}\rangle |m=\pm \tfrac{1}{2} \rangle\,
\end{equation}
where $|\text{BCS}\rangle = |\text{BCS}\rangle_{+}| \text{BCS}\rangle_{-}$, with $|\text{BCS}\rangle_\pm = u_\pm |0\rangle_\pm - v_\pm c^\dagger_{\pm \downarrow} c^\dagger_{\pm \uparrow} |0\rangle_\pm$. The first-order energy correction is given by Eq.~\eqref{eq:two-site-1st-order} in Sec.~\ref{sec:2site}. The second-order correction is defined in Eq.~\eqref{eq:spliteq} and~\eqref{eq:sume}. Using
\begin{equation}
    s^{\pm}_0 |\text{BCS}\rangle =  \mp \alpha |t_{\pm 1}\rangle\,,
\end{equation}
where $\alpha = \frac{1}{2}(u_+v_- - v_+ u_-)$, and $|t_{\pm 1}\rangle$ are the spin-triplet states in Eq.~\eqref{eq:spin-triplets}, we arrive at Eq.~\eqref{eq:splitting-two-site} of the main text.

\section{Details of the NRG calculations}\label{app:nrg}

Considering only the $s$-wave scattering channel, the Hamiltonian in Eq.\eqref{eq:many-body-Hamiltonian} can be represented as a (Wilson chain) 1D lattice using the adaptive scheme introduced in Ref.~\cite{ZITKO20091271_discret} for a constant density of states $\rho(\epsilon)= 1/(2D)$ in the interval $[-D, D]$. The discretization parameter is taken to be $\Lambda=2$. This procedure yields the following Wilson chain Hamiltonian:
\begin{align}
H & =\sum_{i\geq 0} t_i\left[f_{i,\sigma}f^\dag_{i+1,\sigma}+ \mathrm{H.c.} \right] + \Delta\left[f^\dag_{i\uparrow} f^\dag_{i\downarrow} + \mathrm{H.c.}\right] \notag \\ &+h\left[ f_{\uparrow}^\dag(i) f_\uparrow(i)-f_{\downarrow}^\dag(i) f_\downarrow(i)\right]+
J\vec{S}\cdot\vec{s}(0) +\frac{Jh}{2D}S^z ,
\end{align}
where the hopping decays exponentially as $t_N\sim \Lambda^{-N/2}$. Notice that the pairing potential term does not conserve particle number. To render the computation more efficient, we follow the method described in Ref.~\cite{JPSJ.67.1332_BCS_trans} and apply a Bogoliubov transformation:
\begin{align}
b^\dag_{i,\uparrow}&=\frac{1}{\sqrt{2}} \left(f^\dag_{i,\uparrow}+f_{i,\downarrow}  \right),\\
b_{i,\downarrow}&=\frac{1}{\sqrt{2}} \left(f^\dag_{i,\uparrow}- f_{i,\downarrow}  \right).
\end{align}
followed by a particle-hole transformation,
\begin{align}
c^\dag_{2i,\uparrow}&=b^\dag_{2i,\uparrow},\\
c_{2i,\downarrow}&= b_{2i,\downarrow},\\
c^\dag_{2i-1,\uparrow}&=b_{2i-1,\downarrow},\\
c_{2i-1\downarrow}&=-b^\dag_{2i-1,\uparrow}.
\end{align}
Thus, we arrive at the following model:
\begin{align}
H &= \sum_{i\geq 0} \Big\{t_i \sum_{\sigma}\left( c_{i,\sigma} c_{i+1,\sigma}^\dag+ \mathrm{H.c.} \right) \notag\\&+ \Delta   (-1)^i Q^z_i  +2h s^z_i  \Big\}+J \vec{S}\cdot\vec{s}_{0}+\frac{Jh}{2D}S^z.
\end{align}
To reduce the size of the matrices required to diagonalize, the NRG is applied using the following conserved $U(1)$ quantities $(Q^z,S^z_T)$: 
\begin{align}
Q^z&=\sum_{i} Q^z_i = \sum_i \left[ n_{i,\uparrow}+n_{i,\downarrow}-1 \right] ,\\
S^z_T &=S^z + \sum_i s^z(i)= S^z+\frac{1}{2}\sum_i \left[n_{i,\uparrow}-n_{i,\downarrow}\right],
\end{align}
which is suitable for the case with external magnetic fields. At each iteration, we keep at least $1024$ states and discard the states above the energy scale $\omega \approx 10\omega_N=10 \Lambda^{(1-N)/2}$. In the presence of the (superconducting) gap, the NRG iteration must be truncated at iterations with energy scale $\omega_N\ll \Delta$~\cite{Hecht_2008_BCS}. Thus, we stop our NRG computation at iterations with energy scale $\sim 10^{-5}\Delta$ which is sufficient to obtain the spectral properties accurately. We set the temperature $T\ll\Delta$ so effectively that we can consider our results to be in the zero-temperature limit.  For the Kondo model, the spectral weights, $W_\sigma(\epsilon)$, are defined using the T-matrices \cite{PhysRevLett.85.1504_Kondo}. The spectral weight reads,
\begin{align}
W_\sigma(\epsilon) &= -\frac{1}{\pi}\text{Im}\:  C^{R}_{\sigma}(\epsilon),\\
C^{R}_{\sigma}(\epsilon) &= \int dt \, e^{i\epsilon t} C^{R}_{\sigma}(t),\\
C^{R}_{\sigma}(t) &= -i \theta(t) \langle\{ O_{\sigma}(t), O^{\dag}_{\sigma}(0) \} \rangle 
\end{align}
 where $O_\sigma = [ f_{0\sigma}, H_K]$ with $H_K = J \vec{S}\cdot \vec{s}_0$. Note that $f_{0\sigma}$ is the operator in the original basis fermion basis of the superconducting model. To carry out the computation, we obtain the spectral weights using the full density-matrix scheme described in Ref.~\cite{PhysRevLett.99.076402_FDM} and broaden the discrete data set using a hybrid kernel. The spectral function, $A_{\sigma}(\omega)$ is thus computed from the following expression:
\begin{align}
A_\sigma(\omega) &= \sum_\epsilon W_\sigma(\epsilon)\biggl\{\Theta(\epsilon)\Big[\Theta( \epsilon-\epsilon_\text{gap}^{+} ) lG(\omega,\epsilon,a)\notag \\ &+ \Theta( \epsilon_\text{gap}^{+}-\epsilon)G(\omega,\epsilon,b)\Big],\notag
\notag\\
&+\Theta(-\epsilon)\Big[\Theta( \epsilon_\text{gap}^{-} -\epsilon) lG(\omega,\epsilon,a)\notag \\ &+ \Theta(\epsilon- \epsilon_\text{gap}^{-})G(\omega,\epsilon,b)\Big]
\biggr\}.
\end{align}
where
\begin{align}
&lG(\omega,\epsilon,a)\notag\\
&=\frac{\Theta(\omega\epsilon)}{a|\omega|\sqrt{\pi}}\text{Exp}\left[ - \left( \frac{\log(|\omega|)-\log(|\epsilon|)}{a}-\frac{a}{4} \right)^2\right],\\
&G(\omega,\epsilon,b)=\frac{1}{b\sqrt{\pi}}\text{Exp}\left[{-\left(\frac{\epsilon-\omega}{b}\right)^2}\right].
\end{align}
 $\epsilon^+_\text{gap}$ and $\epsilon^-_\text{gap}$ are the positions of the BCS gap at positive and negative sides. They are determined from the data of spectral weights. Outside the gap, we use a logarithmic mesh binning $\sim 500$ points per decade with respect to the gap and a Log-Gaussian kernel with a narrow broadening parameter, $a=0.2$. Inside the gap, we accumulate all the spectral weights and broaden the weights using a Gaussian kernel with width $b=\Delta/1000$. To eliminate the oscillatory artifacts in the continuum due to discretization, the spectral functions are z-averaged~\cite{PhysRevB.41.9403_Z} using $16$ z-points spanning the interval $[1/16,1]$.

\bibliography{NRG}

\end{document}